\documentclass[prb,twocolumn,superscriptaddress,preprintnumbers]{revtex4-2}
\usepackage[utf8]{inputenc}
\usepackage{graphicx}
\usepackage{dcolumn}
\usepackage{bm}
\usepackage{hyperref}
\usepackage{subfigure}
\usepackage{float}
\usepackage{wrapfig}
\usepackage{amssymb,latexsym, amsmath}
\usepackage{graphics}
\usepackage{float}
\usepackage{appendix}
\usepackage{amssymb,latexsym,amsthm,makeidx,enumerate}
\usepackage{mathrsfs}
\usepackage{siunitx}
\usepackage{comment,hhline,multirow,setspace}

\usepackage[table,xcdraw]{xcolor}
\definecolor{navy}{RGB}{0,0,139}
\usepackage{orcidlink}
\usepackage{hyperref}
\hypersetup{colorlinks=true, linkcolor=navy, citecolor=navy, urlcolor=navy, pdftitle={???}, pdfauthor={}}

\newcommand{\NT}{NbTe$_{4}$}
\newcommand{\TT}{TaTe$_{4}$}

\graphicspath{{graphics/}}

\begin{document}
\par\noindent\rule{\textwidth}{0.5pt}
\makeatletter
\let\NAT@bare@aux\NAT@bare
\def\NAT@bare#1(#2){%
	\begingroup\edef\x{\endgroup
		\unexpanded{\NAT@bare@aux#1}(\@firstofone#2)}\x}
\makeatother

\title{Lattice dynamics of the charge density wave compounds TaTe$_4$ and NbTe$_4$ and their evolution across solid solutions}

\author{D. Silvera-Vega}
\affiliation{Department of Physics, \href{https://ror.org/02mhbdp94}{Universidad de Los Andes}, Bogot\'a 111711, Colombia}

\author{G. Cardenas-Chirivi}
\affiliation{Department of Physics, \href{https://ror.org/02mhbdp94}{Universidad de Los Andes}, Bogot\'a 111711, Colombia}

\author{J. A. Galvis \orcidlink{0000-0003-2440-8273}}
\affiliation{School of Sciences and Engineering, Universidad del Rosario, Bogotá 111711, Colombia}

\author{A. C. Garcia-Castro\orcidlink{0000-0003-3379-4495}}
\email{acgarcia@uis.edu.co}
\affiliation{School of Physics, \href{https://ror.org/00xc1d948}{Universidad Industrial de Santander}, SAN-680002, Bucaramanga, Colombia}

\author{P. Giraldo-Gallo\orcidlink{0000-0002-2482-7112}}
\email{pl.giraldo@uniandes.edu.co}
\affiliation{Department of Physics, \href{https://ror.org/02mhbdp94}{Universidad de Los Andes}, Bogot\'a 111711, Colombia}

\date{\today}

\begin{abstract}

Understanding lattice dynamics is central to elucidating the microscopic origin of charge density waves (CDWs), particularly in materials where electron–phonon coupling can play a dominant role. Raman spectroscopy, combined with first-principles calculations, offers a direct means to identify the vibrational modes involved and to monitor their evolution under controlled perturbations. In this work, we combine density functional theory calculations and Raman spectroscopy measurements to investigate the vibrational properties of the quasi-one-dimensional transition metal tetrachalcogenides TaTe$_4$ and NbTe$_4$, as well as their solid solutions Ta$_{1-x}$Nb$_{x}$Te$_4$ (x = 0.0–1.0). 
For the stoichiometric compounds, first-principles calculations predict a phonon instability consistent with the trimerization associated with the CDW phase, providing theoretical evidence for the lattice distortion driving the transition. The calculated Raman-active modes show good agreement with room-temperature experimental spectra, enabling a systematic assignment of the observed peaks. 
Across the solid solution, most Raman modes evolve smoothly with composition. In contrast, the highest-frequency E$_{g}$ mode, dominated by transition-metal motion, exhibits a distinct behavior: its frequency remains close to that of the parent compounds while its intensity redistributes with stoichiometry. This evolution highlights the short-range character of this vibrational mode and suggests its relevance to the CDW-related lattice distortion in these materials.
\\
DOI: XX.XXXX/XxxxXxxX.XXX.XXXXXXX

\end{abstract}

\maketitle

\section{Introduction}
Low-dimensional transition metal chalcogenides have garnered particular attention in recent years \cite{introlibro, IntroLibro2} due to their wide range of emergent electronic phases, including topological states, charge-density waves (CDWs), superconductivity, and ferroic orders \cite{CDWgeneral, Superconductivitygeneral, Magnetismgeneral,PhysRevB.108.155106,doi.org/10.1002/inf2.12156,PhysRevB.102.035125}. 
Within this family, the quasi-one-dimensional transition metal tetrachalcogenides (TMTs), particularly \TT{} and \NT{}, have been proposed to host topologically protected electronic states, as well as CDW order even at room temperature \cite{PhysRevB.102.035125,daschner2025electronholetunnellingprobedhaas}. Therefore, the combination of reduced dimensionality, non-trivial band topology, and symmetry-breaking lattice distortions positions TMTs as model systems for exploring the interplay between topology and collective electronic order.

Recent electronic structure studies indicate that both, \TT{} and \NT{}, may host Dirac and double-Dirac points in their CDW phases \cite{PhysRevB.110.125151,Xu2023,PhysRevLett.133.116403,CDWmodulations}. In addition, both systems become superconductors at very high pressures  \cite{pressuresuperconductivityNbTe4, pressuresuperconductivityTaTe4}. 
Despite these similarities, their transport properties differ significantly: \TT{} exhibits a large residual resistivity ratio (RRR) and pronounced magnetoresistance \cite{10.1063/1.5005907}, whereas \NT{} shows comparatively poor RRR and weaker magnetoresistance \cite{TADAKI1990227}. However, the underlying mechanisms that lead to those discrepancies are not yet fully understood.

In addition to the distinct features observed in the transport properties of \TT{} and \NT{}, distinct features of their CDWs have been reported. Whereas \TT{} has been reported to exhibit a commensurate charge density wave in the full temperature range at and below room temperature \cite{Boswell_1983,BOSWELL198493,https://doi.org/10.1002/pssa.2210770151,mahy1983evidence,FWBoswell_1983}, the nature of the CDW of \NT{} has been the subject of debate. While early X-ray and electron diffraction studies on this last compound showed features consistent with an incommensurate CDW \cite{BOSWELL198493,PhysRevB.34.2979,vanSmaalen:a25451,Böhm+1987+113+122}, recent scanning tunneling microscopy (STM) experiments have revealed that the charge density is locally commensurate in the range of a couple of nanometers, with the presence of CDW phase-slip domain walls between commensurate regions, in agreement with a discommensurate CDW \cite{2023PhRvB.107d5120G}.

Although the details of the CDW modulations in both TMTs have been extensively explored through electronic structure and transport measurements, the microscopic origin of these modulations, as well as the differences between the two systems, remains unsettled. In general, CDW formation in low-dimensional materials can arise from two complementary mechanisms: Fermi surface nesting, where parallel segments of the Fermi surface enhance electronic susceptibility at a specific wavevector, or lattice-driven instabilities due to a soft phonon condensation. In the case of \TT{} and \NT{}, theoretical studies have proposed that the CDW transition is associated with a phonon instability at the CDW wavevector \cite{10.1063/5.0053990,PhysRevB.105.064107}. On the other hand, recent experimental electronic structure investigations, particularly for \TT{}, have not identified clear nesting features capable of stabilizing the observed modulation \cite{Silvera-Vega2026}.

While these results seem to support to a phonon-driven scenario, direct experimental characterization of the lattice dynamics remains limited. A systematic investigation of the phonon spectrum and its evolution is therefore essential to clarify the relative roles of electronic and lattice degrees of freedom in the CDW transition. Such an approach has proven decisive in other CDW materials, including Kagome systems such as KV$_3$Sb$_5$ \cite{luo2022electronic,gutierrez2024phonon} and ScV$_6$Sn$_6$ \cite{hu2024phonon}, where lattice dynamics measurements helped disentangle competing mechanisms.

With this motivation, we present a combined theoretical and experimental investigation of the lattice dynamics in \TT{}, \NT{}, and their solid solutions. Using density functional theory (DFT) calculations and Raman spectroscopy measurements, we systematically characterize the vibrational modes of these compounds. On the theoretical side, we identify a phonon instability consistent with the lattice distortion associated with CDW condensation in both \TT{} and \NT{}. For the stoichiometric compounds, the calculated Raman-active modes show good agreement with the experimental spectra, enabling a consistent indexing and symmetry assignment of the observed peaks. 
Building on this assignment, we investigate the evolution of the Raman modes across the solid solution Ta$_{1-x}$Nb$_x$Te$_4$ ($x$ = 0.0, 0.2, 0.4, 0.6, 0.8, 1.0). We find that the high-frequency $E_g$ modes, dominated by transition-metal motion, retain frequencies close to those of the parent compounds and evolve primarily in intensity with composition, indicating a strong sensitivity to the local metal environment. In contrast, the lower-frequency $E_g$ modes, with a larger Te contribution, shift smoothly with stoichiometry. The remaining Raman-active modes exhibit only minor variations throughout the series.

\begin{figure*}[]
\centering
\includegraphics[width=18.0cm,keepaspectratio=true]{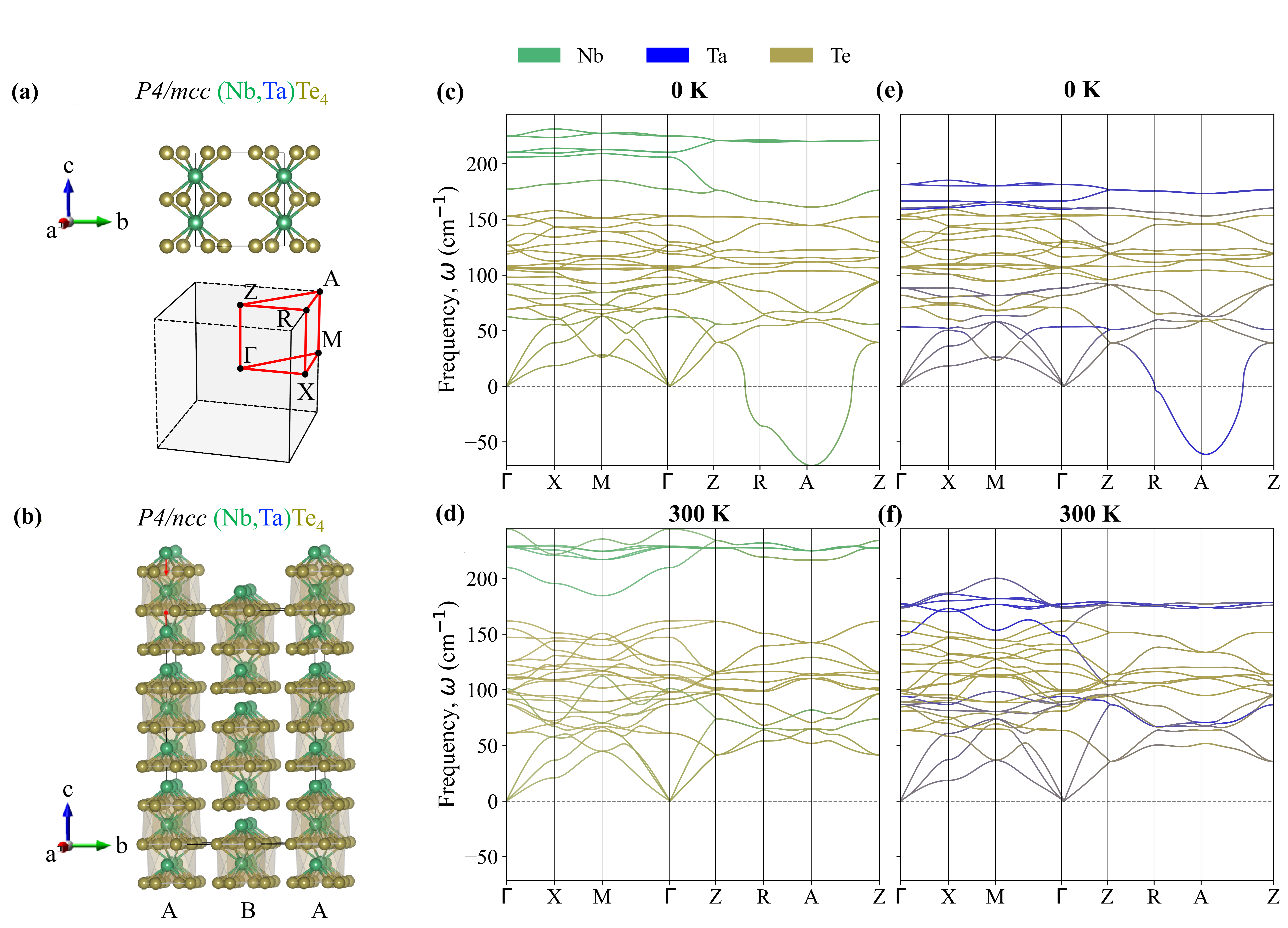}
\caption{(a) (Color online) Crystalline structure of $M$Te$_{4}$, with $M$ = Ta or Nb, alongside their first Brillouin zone in the $P4/mcc$ (SG. 124), or high-temperature non-CDW phase. (b) Crystalline structure of $M$Te$_{4}$ in its CDW phase, which corresponds to the $P4/ncc$ (SG. 130) low-temperature phase. The picture highlights the trimerization of the transition metal along the chains, and the A--B--A stacking of the chains once the commensurate CDW is condensed. (c) Full phonon dispersion curves computed for \NT{} and  (d) and \TT{} at 0 K.  The unstable phonon modes close to the \textit{A} point are denoted at negative frequency values by notation. Similarly, (e) and (f) show the full phonon dispersion curves at 300 K for \NT{} and \TT{}, respectively. Here, the Nb, Ta, and Te sites are denoted in green, blue, and dark olive, respectively, both at the unit cell representations and to indicate phonon bands that are dominated by respective atom motion.}
\label{F1}
\end{figure*}

\section{Experimental and computational methods}

\subsection{Crystal growth and Raman spectroscopy:}
Single crystals of Ta$_{1-x}$Nb$_{x}$Te$_4$ (with $x$= 0.0, 0.2, 0.4, 0.6, 0.8, and 1.0) were grown using the self-flux method. Mixtures of 1 mol\% of Ta$_{1-x}$Nb$_{x}$ and 99 mol\%  high-purity Te were packed into cylindrical alumina crucibles and sealed in evacuated quartz tubes under a pressure of 1.6 mTorr. During synthesis, the temperature was ramped to 700°C over 6 h and kept for 12 h to ensure thorough and homogeneous melting of the compounds. This was followed by a slow cooldown to 500 $^{\circ}$C at a rate of 2.22 $^{\circ}$C/h. The remaining melt was decanted and separated from the Ta$_{1-x}$Nb$_{x}$Te$_4$ crystals using a centrifuge. Silver-colored, long rectangular crystals were obtained, with typical dimensions up to 0.1$\times$0.1$\times$1 cm$^3$.
The crystallinity of the phases was analyzed by X-ray  powder diffraction employing a Panalytical Empyrean diffractrometer equipped with a Cu-K$_\alpha$ source.

Raman measurements were performed at room temperature using a Horiba XPlora spectrometer. A grating resolution of 2400 lines/mm was used for the 532 nm and 638 nm laser excitations, while a resolution of 1800 lines/mm was used for the 738 nm laser.

\subsection{Computational details and theoretical approaches:}

First-principles calculations were performed within the density functional theory (DFT) \cite{PhysRev.136.B864,PhysRev.140.A1133} approach using the \textsc{vasp} code (version 5.4.4) \cite{PhysRevB.54.11169,PhysRevB.59.1758}. 
The projector augmented-wave (PAW) method \cite{PhysRevB.50.17953}, represented the valence and core electrons. 
The valence electronic configurations considered in the pseudopotentials are Nb: (4$s^2$4$p^6$5$p^1$4$d^4$, version 25May2007),  Ta: (6$s^2$5$d^3$, version 17Jan2003), and Te: (5$d^2$5$p^4$, version 08Apr2002). 
The exchange-correlation functional was treated using the GGA-PBEsol parameterization of the generalized gradient approximation \cite{PhysRevLett.100.136406}. 
The periodic crystal was modeled using Bloch states with a Monkhorst-Pack \cite{PhysRevB.13.5188} \emph{k}-point mesh of 13$\times$13$\times$11 and 600 eV energy cutoff to give a force convergence of less than 0.001 eV$\cdot$\r{A}$^{-1}$ and an error in energy less than ~10$^{-6}$ eV.  
The phonon calculations were performed within the finite-differences methodology \cite{Monserrat_2018} and analyzed through the \textsc{Phonopy} interface \cite{phonopy-phono3py-JPCM}. 
The latter calculations were performed in a (2$\times$2$\times$2) supercell to properly map the lattice dynamics at the zone boundary.
To account for the anharmonic interaction, molecular dynamics simulations were performed at 300 K. Interatomic forces were postprocessed using the \textsc{DynaPhopy} \cite{Carreras_2017}.
Atomic structure visualizations were created using the \textsc{VESTA} software \cite{momma_vesta_2011}. \cite{HERATH2020107080, LANG2024109063}.

\section{Results and discussion}

Both TMTs, \TT{} and \NT{}, crystallize at high temperature in a tetragonal unit cell under the $P4/mcc$ space group (No. 124), and are well-known quasi-one-dimensional (quasi-1D) materials \cite{10.1063/1.5005907,TADAKI1990227}. 
In their stoichiometric form MX$_4$, these compounds are composed of chains of transition metal atoms (M = Ta or Nb) located at the center of two Te-square antiprisms (X = Te), which extend along the crystallographic \textit{c}-axis, see Fig. \ref{F1}(a). 
These chains are coupled to neighboring chains through van der Waals forces, which explains the quasi-1D nature of TMTs. The $P4/mcc$ structure corresponds to the non-CDW phase, whose unit cell is shown within the small rectangular prism in Fig. \ref{F1}(a), along with its corresponding first Brillouin zone, where high-symmetry points are labeled in capital letters. 

Once the temperature is lowered below the transition temperature of $T \approx 475$ K in TaTe$_4$ \cite{10.1063/1.4977708}, a displacive phase transition occurs from the $P4/mcc$ to the $P4/ncc$ space group leading to a commensurate CDW phase. In contrast, the  transition temperature for NbTe$_4$ is not as well-established \cite{TADAKI1990227,Böhm1999}, and there are reports of incommensurate and discommensurate CDW phases \cite{Böhm1999,2023PhRvB.107d5120G}. While the CDW in TaTe$_4$ and NbTe$_4$ display different features, the CDW transition for both TMT representatives can be explained from the trimerization of the M sites in the one-dimensional chains \cite{PhysRevB.107.045120}. In the commensurate CDW structure, the unit cell expands to $2a \times 2a \times 3c$ and is denoted by the large rectangular prism in Fig. \ref{F1}(b). The doubling of the $a$ parameter in the superstructure is a consequence of the A–-B–-A stacking of trimers in adjacent chains. Such structural stacking leads to a shift between trimers in neighboring chains A and B by one M atom (i.e., $c/2$).

According to the symmetry operations of space group No. 124 ($P4/mcc$), and considering that the $M$ site occupies the 2$a$ Wyckoff position while the Te sites are located at the 8$m$ positions, the irreducible representation of the MTe$_4$ unit cell is given by:
${\Gamma} = 2A_{1g} \oplus A_{1u} \oplus 3A_{2g} \oplus 2A_{2u} \oplus 2B_{1g} \oplus B_{1u} \oplus 2B_{2g} \oplus B_{2u} \oplus 5E_u \oplus 3E_g.$

\begin{table*}[t]
\centering
\caption{Computed IR and Raman modes at 0 K in the pristine NbTe$_4$ and TaTe$_4$ with irreducible representations $\Gamma = 2A_{1g} \oplus A_{1u} \oplus 3A_{2g} \oplus 2A_{2u} \oplus 2B_{1g} \oplus B_{1u} \oplus 2B_{2g} \oplus B_{2u} \oplus 5E_u \oplus 3E_g$. Additionally, the experimentally Raman measured frequencies $\omega_{exp}$ are also included for comparison.}
\begin{tabular*}{\textwidth}{@{\extracolsep{\fill}} l c c | l c }
\hline
\hline
NbTe$_4$ \rule[-1ex]{0pt}{3.5ex}  \\
\hline
Raman Mode  &  $\omega$ (cm$^{-1}$) & $\omega_{exp}$ (cm$^{-1}$) & IR-Mode &   $\omega$ (cm$^{-1}$)   \rule[-1ex]{0pt}{3.5ex} \\
\hline
$A_{1g}$ &  129.6, 144.7 & 137.4, 144.3 & $A_{1u}$ & 121.2   \rule[-1ex]{0pt}{3.5ex}  \\
$B_{1g}$ &  106.6, 153.2 & 120.9, --- & $A_{2g}$ & 62.5, 118.9, 177.1   \rule[-1ex]{0pt}{3.5ex}  \\
$B_{2g}$ &  108.0, 152.6 & 121.4, --- & $A_{2u}$ & 205.8   \rule[-1ex]{0pt}{3.5ex}  \\
$E_{g}$ &  91.7, 106.0, 210.3& 87.2, 98.9, 213.8 & $B_{1u}$& 98.0   \rule[-1ex]{0pt}{3.5ex}  \\
---            &  ---  &  & $B_{2u}$& 104.6   \rule[-1ex]{0pt}{3.5ex}  \\
---            &  ---  &  & $E_{u}$& 69.1, 82.3, 126.8, 224.8
\rule[-1ex]{0pt}{3.5ex}  \\
\hline
TaTe$_4$ \rule[-1ex]{0pt}{3.5ex}  \\
\hline
Raman Mode  &  $\omega$ (cm$^{-1}$) & $\omega_{exp}$ (cm$^{-1}$) & IR-Mode &   $\omega$ (cm$^{-1}$)   \rule[-1ex]{0pt}{3.5ex} \\
\hline
$A_{1g}$ & 131.0, 150.2 & 136.0, 145.8 & $A_{1u}$ & 124.7 \rule[-1ex]{0pt}{3.5ex} \\
$B_{1g}$ & 106.4, 154.4 & 125.5, --- & $A_{2g}$ & 53.4, 116.2, 160.0 \rule[-1ex]{0pt}{3.5ex} \\
$B_{2g}$ & 107.7, 153.1 & 120.4, --- & $A_{2u}$ & 158.9 \rule[-1ex]{0pt}{3.5ex} \\
$E_{g}$ & 88.2, 107.0, 166.6 & 80.4, 93.1, 174.6 & $B_{1u}$ & 101.5 \rule[-1ex]{0pt}{3.5ex} \\
---            &  ---  &  & $B_{2u}$& 107.3 \rule[-1ex]{0pt}{3.5ex} \\
---            &  ---  &  & $E_{u}$& 70.6, 81.7, 129.6, 181.2
\rule[-1ex]{0pt}{3.5ex}  \\
\hline
\hline
\end{tabular*}
\label{tab:modes}
\end{table*}

At the initial stage, and as shown in Figs. \ref{F1}(c) and \ref{F1}(e), our quasi-harmonic phonon calculations at 0 K reveal the phonon dispersion relations of both TMTs. In both cases, an imaginary frequency, indicating a vibrational instability, is predicted near the $A$ point, with a minimum close to $(\frac{1}{2}a, \frac{1}{2}a, \frac{1}{3}c)$. This instability is displayed using negative frequency values by convention, and is consistent with previous calculations in this compound \cite{10.1063/5.0053990,PhysRevB.105.064107}. For NbTe$_4$, the computed unstable mode occurs at $\omega = -71.3$ cm$^{-1}$, while for TaTe$_4$, it appears at $\omega = -61.2$ cm$^{-1}$. This phonon is consistent with a symmetry allowed phase transition, an $E_g$ mode. This instability is a typical signature of a Kohn anomaly, an essential ingredient for CDW formation \cite{Gruner1994}.
Interestingly, no other unstable mode is observed. Moreover, in the atomically-projected phonon dispersion curves, displayed in Figs. \ref{F1}(c) and \ref{F1}(e) it can be observed that this mode is mainly M driven.

The atomic displacements associated with this instability involve intrachain motion along the $c$-axis of the transition metal in both TMT representatives. Such displacements closely resemble the trimerization of the $M$ atoms and the $A-B-A$ stacking pattern observed experimentally in these compounds, as revealed by scanning tunneling microscopy (STM) \cite{2023PhRvB.107d5120G}.

Going further and aiming to theoretically estimate the anharmonic distortion and phonon spectrum once the unstable phonon mode is allowed to condense, we performed first-principles-based molecular dynamics calculations at room temperature.  At 300 K, the main effect of molecular dynamics on the phonon dispersion of both TMT compounds is the complete removal of the previously imaginary ($i.e.$ unstable) soft phonon, as shown in Figs. \ref{F1} (d) and (f).  This difference arises from the anharmonic interactions naturally included in the finite-temperature molecular dynamics simulations, which renormalize the phonon spectrum and suppress the lattice instability present at zero temperature. Apart from that, a similar set of modes can be obtained for both phases. The entire list of computed IR and Raman modes is presented in Table \ref{tab:modes}.

\begin{figure*}[]
\centering
\includegraphics[width=18.0cm,keepaspectratio=true]{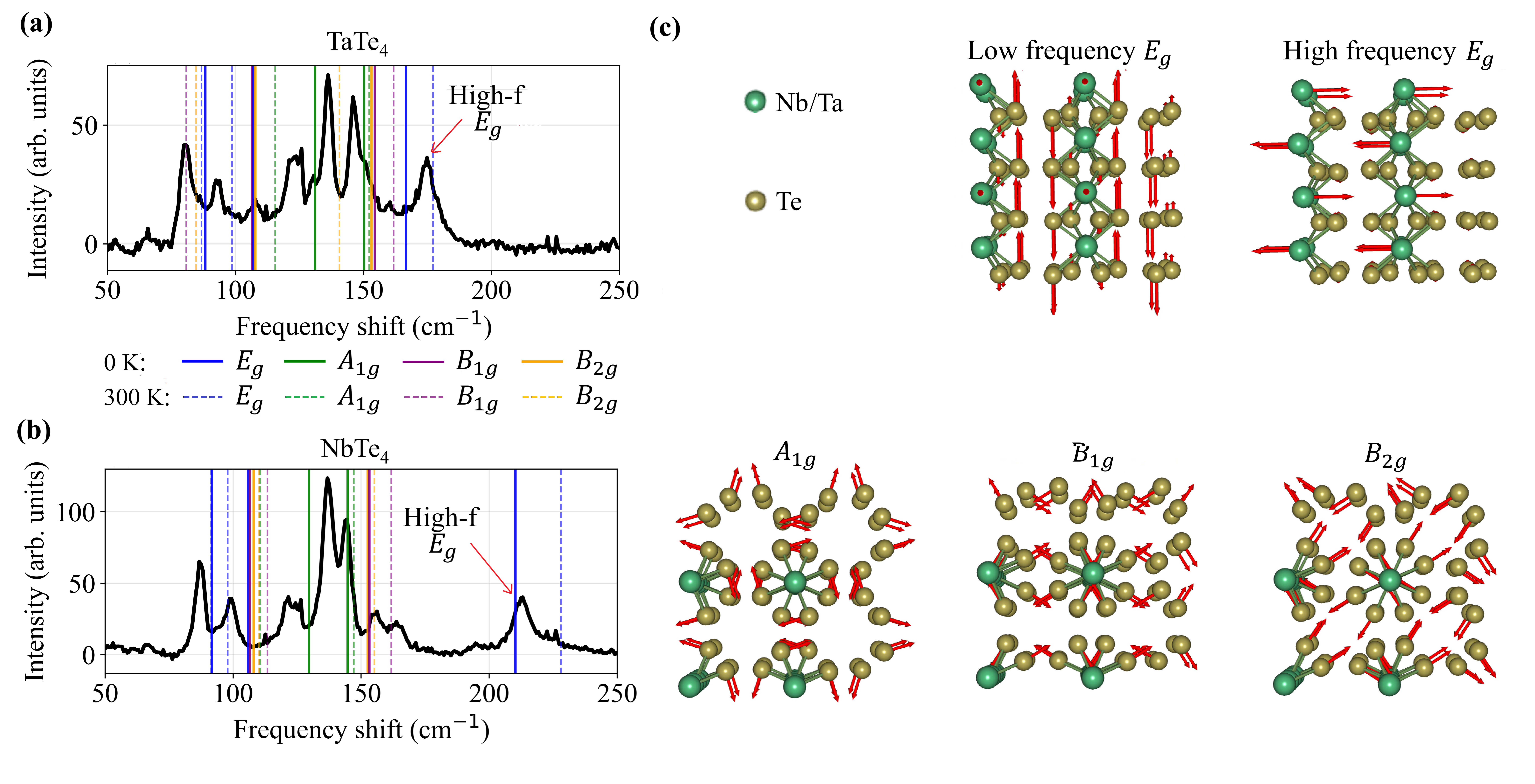}
\caption{(Color online) (a) Raman spectroscopy measurements performed at room temperature for \NT{} and (b) \TT{}, and their comparison with theoretical predictions at both 0 K and 300 K. Solid (dashed) vertical lines represent the calculated positions for the 0 K (300 K) modes. The different colors used represent the different irreps at the $\Gamma$ high-symmetry point of each of the predicted modes: blue for $E_g$, green for $A_{1g}$, purple for $B_{1g}$ and orange for $B_{2g}$ modes. (c) Schematic of atomic motion (red arrows) associated to selected vibrational modes in \TT{} and \NT{}, labeled according to the irreducible representation of the high-temperature groups at the $\Gamma$ high symmetry points. The low frequency $E_g$ mode displayed here is the one appearing at approximately 80.4 cm$^{-1}$ for TaTe$_4$ and 87.2 cm$^{-1}$ in the Raman data. The high frequency $E_g$ mode appears at 174.6 cm$^{-1}$ in \TT{} and at 213.8 cm$^{-1}$ in \NT{}. The presented $A_{1g}$ mode appears at 136.0 (137.4) and 145.8 (144.3) cm $^{-1}$ for \TT{} (\NT{}),  while $B_{1g}$ and $B_{2g}$ modes appear at 125.5 (124.1) and 120.4 (120.9) cm$^{-1}$. }
\label{fig:deconvolution}
\end{figure*}

In order to gain further insights and correlate the theoretical results obtained so far, we performed an experimental characterization of the vibrational spectrum of TMT single crystals via Raman spectroscopy. The positions of the experimental peaks obtained from Raman measurements were compared with vibrational mode frequencies and their associated irreducible representations (irreps), calculated using density functional theory (DFT) at both 0 K and 300 K. These comparisons are shown in Figs. \ref{fig:deconvolution}(a) and \ref{fig:deconvolution}(b), corresponding to the pure \TT{} and \NT{} samples, respectively.
Although the room temperature phase of these compounds belongs to the distorted P4/ncc space group, the mode assignment was performed using the phonon spectrum of the high-symmetry  P4/mcc structure. This choice allows the vibrational modes to be described within the primitive cell of the undistorted phase, avoiding additional band folding effects introduced by the CDW superstructure. In this unfolded representation, each mode can be more directly compared with the experimentally observed Raman peaks, whereas a folded description would artificially increase the apparent number of branches without introducing new independent vibrational degrees of freedom.

From a theoretical standpoint, temperature primarily causes a shift of the predicted peaks toward lower wavenumbers due to anharmonic lattice dynamics. However, the number of Raman-active modes remains the same at both temperatures, since the same primitive cell is considered in both cases. Therefore, experimental peak assignments can be reliably made using either set of calculated irreps, accounting only for the small distortions induced by the CDW.  
Interestingly, the mode frequencies associated with the 0 K irreps showed closer agreement with the experimental data, despite measurements being taken at room temperature. The better agreement obtained with the 0 K results suggests that the finite-temperature approach may slightly overestimate the magnitude of mode softening \cite{luo2022electronic,Togo_2023, PhysRevB.92.054301}.

\begin{figure*}[!t]
        \centering
        \includegraphics[width=18.0cm,keepaspectratio=true]{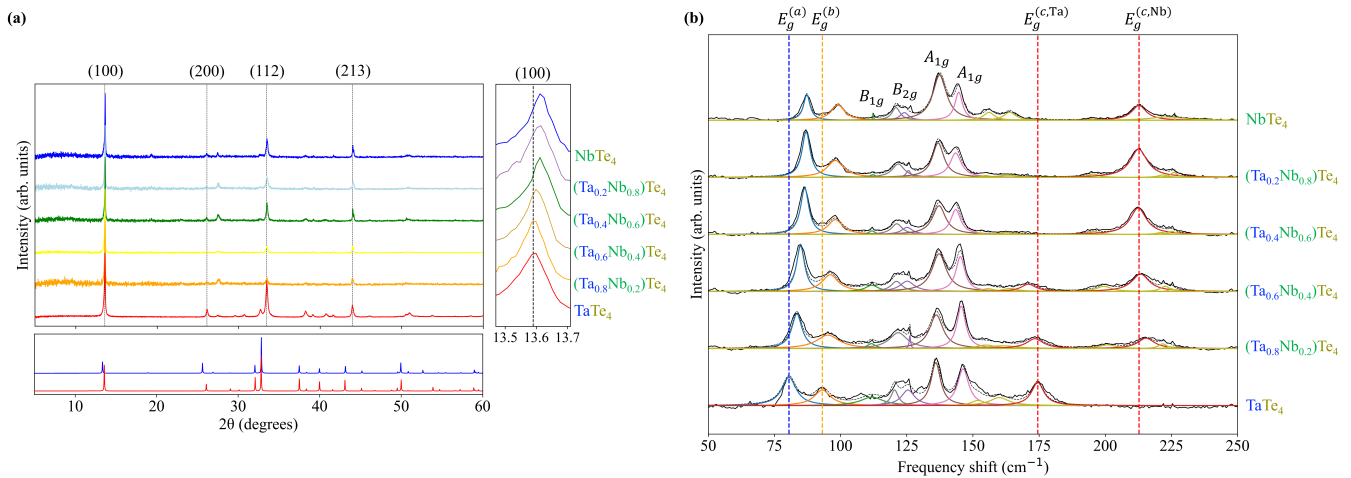}
        \caption{ (Color online) (a) Measured XRD patterns for the entire family of Ta$_{1-x}$Nb$_x$Te$_4$ crystals. The curve at the bottom of the upper panel corresponds to x = 0
(pure \TT{}) and the one at the top corresponds to x = 1 (pure \NT{}). The curves in between correspond to the doping series. The right panel shows a zoom-in to the (100) peaks. The lower panel shows the predicted XRD pattern for \NT{} (blue) and \TT{} (red), as calculated from the CIF file in the Materials Project website \cite{Horton2025}. (b) Evolution of Raman peaks as a function of Nb doping rate $x$ across the solid solution Ta$_{1-x}$Nb$_{x}$Te$_{4}$. Here, the curve at the bottom corresponds to $x = 0$ (pure \TT{}), the one at the top corresponds to $x = 1$ (pure \NT{}), and intermediate compositions in between. The individual peaks for each $x$ value were fitted to individual Lorentzians, and presented in different colors according to the irrep assignment of Fig. 2(a,b). The black dashed curves correspond to the sum of the Lorentzian fits for each $x$, and the solid black lines correspond to the actual experimental data. The dashed vertical lines are used as visual guides to highlight the evolution of the different $E_g$ peaks. }
        \label{fig:cascade}
    \end{figure*}

Although both \NT{} and \TT{} exhibit the same number of Raman-active modes according to DFT, with the same characteristic vibrations (some of which are illustrated in Fig. \ref{fig:deconvolution}(c)), there are notable differences in the positions of these peaks. The most significant of these differences concerns the highest-frequency $E_g$ mode in each material. According to theoretical predictions, the highest $E_g$ mode in \NT{} appears near 210 cm$^{-1}$, whereas in \TT{} the corresponding $E_g$ mode is located around 166 cm$^{-1}$, as shown in Figs. \ref{fig:deconvolution}(a) and \ref{fig:deconvolution}(b). 
This marked discrepancy is also observed in the experimental spectra, although the measured peaks are slightly shifted relative to the theoretical values. This behavior contrasts with other irreducible representations, for which the predicted mode frequencies in both \NT{} and \TT{} are relatively similar, including other lower frequency $E_g$ modes, for which there is a shift in frequency when comparing these compounds, but to a lesser extent.  This suggests that although these high and low frequency $E_g$ modes share the same symmetry label, the atomic displacements differ significantly, an aspect that will be explored in more detail later.  

While there are differences in the distinct $E_g$ vibrations in TMTs, there is an overall softening in these particular modes for \TT{} when comparing with \NT{}. This behavior can be understood in terms of the dynamical matrix that governs the vibrational modes. The fact that the $E_g$ mode frequencies in \TT{} are lower than those in \NT{} suggests that the diagonal elements of the dynamical matrix—associated with atomic masses—are significantly affected, as Ta has nearly twice the mass of Nb. This evolution is in fair agreement with the behavior expected for a simple harmonic oscillator, for which $\omega=\sqrt{k/m}$.

In contrast to recent Raman studies that reported a drastic difference in the Raman peaks intensity between stoichiometric \TT{} and \NT{} \cite{https://doi.org/10.1002/jrs.6661}, our experimental data do not exhibit such effects. While those prior observations were attributed to differences in the CDW condensate between the two materials \cite{https://doi.org/10.1002/jrs.6661}, such a conclusion cannot be drawn from our measurements. Moreover, our data reflects close agreement with theoretical Raman peaks, and we do not see any evidence of relaxation of Raman selection rules due to the CDW condensation that might give rise to observation of forbidden or infrared modes, as  suggested by Z. E. Nataj et al. \cite{https://doi.org/10.1002/jrs.6661}.

\begin{figure*}[]
        \centering
        \includegraphics[width=18.0cm,keepaspectratio=true]{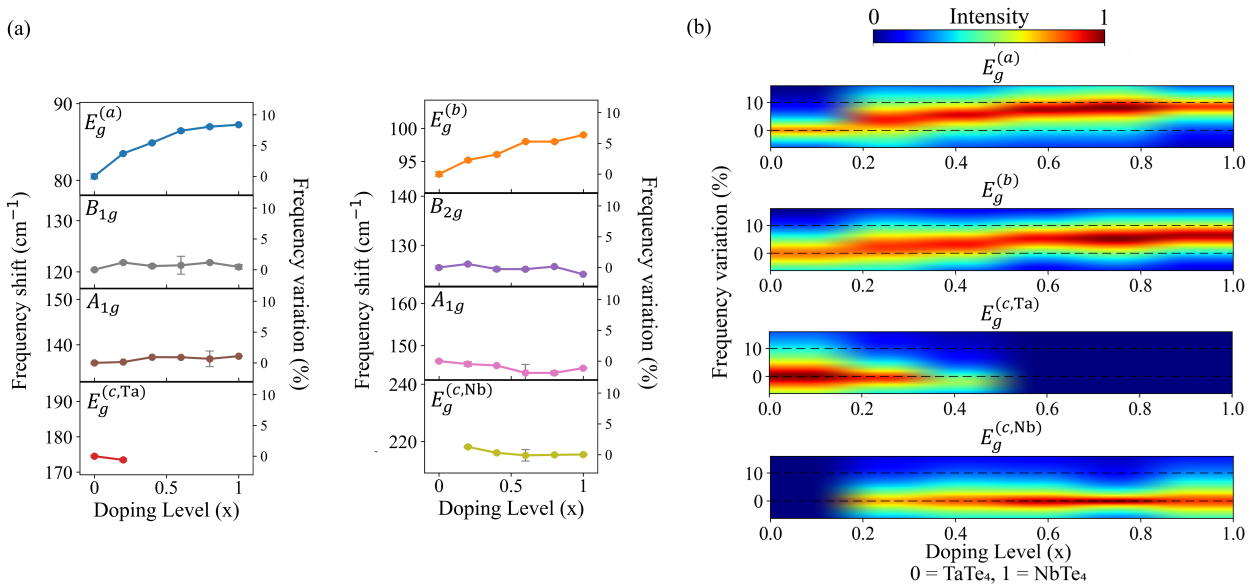}
        \caption{(Color online) : (a) Frequencies of the identified Raman peaks as a function of chalcogen doping. (b) Heatmaps showing the percentage variation of the different $E_g$ peaks as a function of doping level $x$, where $x=0$ for \TT{} and $x=1$ for \NT{}. Here, the color indicates the intensity of the peak for each doping level. Blue represents peak absence, while intense red indicates doping levels where the intensity of the peak is the highest. The horizontal dashed lines are a visual guide to highlight the relative percentage evolution of the $E_g$ modes throughout the whole doping series.}
        \label{fig:cascade2}
    \end{figure*}

To systematically track the evolution of lattice dynamics features of TMTs as going from \NT{} to \TT{}, single crystals of the solid solution Ta$_{1-x}$Nb$_{x}$Te$_4$, where $x$ denotes the Nb concentration relative to Ta, were characterized. X-ray diffraction (XRD) data for the doping series are shown in Fig. \ref{fig:cascade}(a). The XRD patterns show that the diffraction peaks gradually shift to larger 2$\theta$ values as the doping series evolves from TaTe$_4$ to NbTe$_4$, indicating a contracting unit cell, as expected. This evolution is better observed in the close-up look to the (100) peak, in the right in fig. 3(a). From the observed peaks, the lattice parameters $(a,c)=(6.510,6.810)$ \r{A} and  $(a,c)=(6.494,6.792)$ \r{A} were obtained for TaTe$_4$ and NbTe$_4$, respectively, with good agreement with the values reported in the literature \cite{FWBoswell_1983,Silvera-Vega2026,2023PhRvB.107d5120G}. From the angle shift of the peaks, a continuous evolution of lattice parameter values for intermediate compositions in the solid solutions is deduced, which indicates a homogeneous Nb/Ta doping for the solid solution crystals.

Raman spectroscopy measurements were also performed in the Ta$_{1-x}$Nb$_{x}$Te$_4$ samples, as presented in Fig. \ref{fig:cascade}(b). All compositions show Raman peak intensities comparable to the intensities of the corresponding peaks of the parent compounds, and a smooth evolution of their peak frequencies across the doping series. A spectral deconvolution procedure was applied using individual Lorentzian fits to isolate the information of each vibrational mode for each composition. This approach allowed us to label the experimental peaks by comparing them with DFT-predicted Raman-active modes. Each Lorentzian component was colored consistently to facilitate visual tracking of the peak evolution across the doping series.  An interesting feature for intermediate compositions of the doping series is the presence of an additional peak in the high frequency range. By carefully observing the evolution across the doping series, it can be seen that the two highest frequency peaks for intermediate concentrations are in close proximity to the highest frequency $E_g$ mode of the pure compounds, labeled as $E_g^{(c,\text{Ta})}$ and $E_g^{(c,\text{Nb})}$ and highlighted by dashed vertical lines in Fig. \ref{fig:cascade}(b). For these two peaks, in the intermediate compositions of the doping series, their relative intensities vary proportional to their corresponding metal ion concentration. In contrast, the lower frequency $E_g$ modes, labeled as $E_g^{(a)}$ and $E_g^{(b)}$ and also marked with vertical dashed lines at the positions corresponding to the $x=0$ frecuency values in Fig. \ref{fig:cascade}(b), evolve smoothly in frequency maintaining their intensities across the doping series. These observations are analyzed in Fig. \ref{fig:cascade2}, and will be discussed in detail next. 

After the deconvolution analysis, we plotted the evolution of the frequency of the different identified modes as a function of Nb doping, as shown in Fig. \ref{fig:cascade2}(a). In order to highlight the peak intensity variations, the heatmaps in Fig. \ref{fig:cascade2}(b) present in color the variations in intensity for the $E_g$ peaks, each normalized to the intensity of their corresponding lowest frequency $A_{1g}$ peaks (as this is in general the most intense for all compositions), and in the vertical scale the percentage variations in frequency with respect to the $x=0$ value, as a function of doping. For the $E_g^{(c,\text{Nb})}$ the percentage variation in frequency was calculated with respect to the $x=1$ value, as this peak is not present for pure \TT{}.

We first discuss the low-frequency $E_g$ modes ($E_g^{(a)}$ and $E_g^{(b)}$ in Fig \ref{fig:cascade2}(a,b)). These modes progressively shift to higher wavenumbers as going from \TT{} to \NT{}, with a total variation of about 10$\%$ across the series, while their peak intensity remains relatively constant across the series (evidenced in a sustained  red color in the heatmap of fig. \ref{fig:cascade}(b)). To understand this behavior, it is necessary to consider the nature of these modes beyond their symmetry labels. Although all $E_g$ modes involve intrachain vibrations, $E_g^{(a)}$ and $E_g^{(b)}$ modes are primarily driven by Te atom motion along the $c$-axis, while the M atoms undergo very slight displacements along the $a$-axis, as illustrated in Fig. \ref{fig:deconvolution}(c). Such distortion results in a nearly rigid motion of atoms along the chain and minimally alters the arrangement of M-trimer clusters that define the CDW phase. Consequently, these modes are less sensitive to the individual identity of the transition metal and display a smooth and monotonic evolution with doping, probing the macroscopic average value of transition metal content.

In contrast, the high-frequency $E_g$ modes ($E_g^{(c,\text{Ta})}$ and $E_g^{(c,\text{Nb})}$ in Fig \ref{fig:cascade}(b)) exhibit a markedly different behavior. At intermediate doping levels, two distinguishable frequencies can be associated with the same symmetry channel, one at about 170 cm$^{-1}$ and the other at approximately 213 cm$^{-1}$, which can be associated to $E_g^{(c,\text{Ta})}$ and $E_g^{(c,\text{Nb})}$, respectively, as already discussed. These two peaks can be traced across different dopings, but their intensity varies depending on which transition metal site dominates the chemical structure. The highest-frequency peak, associated to $E_g^{(c,\text{Nb})}$, only appears if there is Nb in the solid solution (for $x>0.2$, as seen in the heatmap), and its intensity increases as Nb ($x$) content increases. Interestingly, its frequency value remains pinned around the value seen in pure \NT{}. The other high-frequency $E_g$ peak ($E_g^{(c,\text{Ta})}$) appears at stoichiometric \TT{} and its amplitude dampens as Nb content increases, completely vanishing for compositions $x>0.6$, as approaching pure \NT{}. Its frequency value also remains pinned, but around the value seen in pure \TT{}, as highlighted in Figs. \ref{fig:cascade2}(a,b).

The contrasting evolution between the different $E_g$ peaks can be understood by considering the character of their atomic displacements. Unlike the low-frequency $E_g$ modes $E_g^{(a)}$ and $E_g^{(b)}$, the high-frequency $E_g$ modes $E_g^{(c,\text{Ta})}$ and $E_g^{(c,\text{Nb})}$ are predominantly M-driven, as evidenced by the projected atomic contributions in the phonon band dispersions shown in Figs. \ref{F1}(c)–(f). In these modes, the dominant motion of the transition metal atoms is confined to the $aa$-plane, as illustrated in Fig. \ref{fig:deconvolution}(c). Neighboring M atoms along the chain oscillate out of phase, dynamically modulating the M–Te bond lengths and directly perturbing the trimer configuration associated with the CDW state. 
The strong metal character of these vibrations explains why their frequencies do not evolve smoothly between the \TT{} and \NT{} limits. Instead, each mode remains pinned near the frequency of its respective parent compound, with its intensity varying proportionally to the Ta/Nb content. Given the smooth evolution of all other Raman modes and the continuous change of lattice parameters observed in XRD, a macroscopic phase separation between \TT{} and \NT{} can be excluded. Rather, the anomalous behavior of the high-frequency $E_g^{(c)}$ modes points to a short-range character, reflecting the immediate local environment of the transition metal site. 
Importantly, because these modes involve direct modulation of the metal sublattice and can strongly distort the CDW-related trimer structure, their unusual evolution may be connected to the distinct CDW characteristics reported for \TT{} and \NT{}. While a definitive causal link cannot yet be established, it is plausible that this high-frequency $E_g$ vibration encodes key information about the differing stability, commensurability, or domain structure of the CDW states in the two compounds. In this sense, this mode may provide a particularly sensitive probe of the microscopic mechanisms underlying the contrasting CDW behavior in these closely related materials.

In contrast to the $E_g$ modes, other Raman-active modes remain relatively stable across the solid solution, as shown in Figs. \ref{fig:cascade2}(a), for the $A_{1g}$, $B_{1g}$, and $B_{2g}$ modes. This behavior stems from the nature of the corresponding atomic displacements. While the $E_g$ modes involve intrachain motion, the $A_{1g}$, $B_{1g}$, and $B_{2g}$ modes are predominantly Te-driven and can be described as interchain vibrations. As a result, the frequencies of these modes are expected to evolve only minimally with doping, since their dynamics are not as sensitive to the metal content.

\section{Conclusions and general remarks}

In this work, we presented a comprehensive theoretical and experimental investigation of the lattice dynamics in \TT{}, \NT{}, and their solid solutions Ta$_{1-x}$Nb$_x$Te$_4$. By combining density functional theory calculations with Raman spectroscopy measurements, we achieved a consistent assignment of the Raman-active modes in the stoichiometric compounds and established good agreement between calculated and measured vibrational frequencies. Theoretical calculations reveal a phonon instability at the CDW wavevector in both parent materials, consistent with the trimerization of the transition-metal chains and supporting a lattice-driven component in the CDW formation.
Extending the analysis across the solid solution enabled us to track the evolution of the vibrational spectrum under controlled modification of the transition-metal sublattice. While most Raman-active modes evolve smoothly with composition, the highest-frequency $E_g$ mode, dominated by transition-metal motion, exhibits an anomalous behavior: its frequency remains pinned near that of the respective parent compound while its intensity redistributes with stoichiometry. This distinct evolution, in contrast to the monotonic behavior of Te-dominated modes, indicates a strong sensitivity of this vibration to the local metal environment and highlights its short-range character.
Because this high-frequency $E_g$ mode directly modulates the metal sublattice and perturbs the CDW-related trimer configuration, its unusual behavior may be relevant to understanding the distinct CDW properties reported for \TT{} and \NT{}. Although further investigations are required to establish a direct connection, our results suggest that this vibrational mode provides a particularly sensitive probe of the interplay between lattice dynamics and CDW modulation in these quasi-one-dimensional materials. More broadly, this study demonstrates that systematic phonon characterization across solid solutions offers valuable insight into the microscopic mechanisms governing CDW formation in low-dimensional systems.

\section*{Acknowledgements}
The authors acknowledge Nathalia Cardona Cañavera and Nathalia Vera Gallo for support on the data taking of the Raman spectra of some samples, and the support of the X-ray diffraction and fluorescence laboratory of the Faculty of Science of Universidad de Los Andes. The authors acknowledge the instruments and scientific and technical assistance of the MicroCore Microscopy Core at the Universidad de Los Andes, a facility that is supported by the vicepresidency for research and creation. 
A.C.G.C. acknowledges the computational support of Laboratorio de Supercomputo del Sureste (LNS), Benemérita Universidad Autónoma de Puebla, BUAP, extended to us for performing heavy theoretical calculations.
A.C.G.C. acknowledges the grant No. 4211 entitled “Búsqueda y estudio de nuevos compuestos antiperovskitas laminares con respuesta termoeléctrica mejorada para su uso en nuevas energías limpias” supported by Vicerrectoría de Investigaciones y Extensión, VIE-UIS. 
P.G-G. gratefully acknowledges funding from the project ‘Ampliación del uso de la mecánica cuántica desde el punto de vista experimental y su relación con la teoría, generando desarrollos en tecnologías cuánticas útiles para metrología y computación cuántica a nivel nacional’, BPIN 2022000100133, from SGR of MINCIENCIAS, Gobierno de Colombia.
P.G-G. also acknowledges funding from the project No. 5.187 of the 'Fundación para promoción de la investigación y la tecnología' of the Banco de la República de Colombia. 

\section*{Data availability:}
Data used and produced in the development of this work will be made available from the author upon reasonable request.

\bibliography{1_bib_NbTe4.bib}

\providecommand{\noopsort}[1]{}\providecommand{\singleletter}[1]{#1}%
\begin{thebibliography}{53}%
\makeatletter
\providecommand \@ifxundefined [1]{%
 \@ifx{#1\undefined}
}%
\providecommand \@ifnum [1]{%
 \ifnum #1\expandafter \@firstoftwo
 \else \expandafter \@secondoftwo
 \fi
}%
\providecommand \@ifx [1]{%
 \ifx #1\expandafter \@firstoftwo
 \else \expandafter \@secondoftwo
 \fi
}%
\providecommand \natexlab [1]{#1}%
\providecommand \enquote  [1]{``#1''}%
\providecommand \bibnamefont  [1]{#1}%
\providecommand \bibfnamefont [1]{#1}%
\providecommand \citenamefont [1]{#1}%
\providecommand \href@noop [0]{\@secondoftwo}%
\providecommand \href [0]{\begingroup \@sanitize@url \@href}%
\providecommand \@href[1]{\@@startlink{#1}\@@href}%
\providecommand \@@href[1]{\endgroup#1\@@endlink}%
\providecommand \@sanitize@url [0]{\catcode `\\12\catcode `\$12\catcode `\&12\catcode `\#12\catcode `\^12\catcode `\_12\catcode `\%12\relax}%
\providecommand \@@startlink[1]{}%
\providecommand \@@endlink[0]{}%
\providecommand \url  [0]{\begingroup\@sanitize@url \@url }%
\providecommand \@url [1]{\endgroup\@href {#1}{\urlprefix }}%
\providecommand \urlprefix  [0]{URL }%
\providecommand \Eprint [0]{\href }%
\providecommand \doibase [0]{https://doi.org/}%
\providecommand \selectlanguage [0]{\@gobble}%
\providecommand \bibinfo  [0]{\@secondoftwo}%
\providecommand \bibfield  [0]{\@secondoftwo}%
\providecommand \translation [1]{[#1]}%
\providecommand \BibitemOpen [0]{}%
\providecommand \bibitemStop [0]{}%
\providecommand \bibitemNoStop [0]{.\EOS\space}%
\providecommand \EOS [0]{\spacefactor3000\relax}%
\providecommand \BibitemShut  [1]{\csname bibitem#1\endcsname}%
\let\auto@bib@innerbib\@empty
\bibitem [{\citenamefont {Meerschaut}\ and\ \citenamefont {Rouxel}(1986)}]{introlibro}%
  \BibitemOpen
  \bibfield  {author} {\bibinfo {author} {\bibfnamefont {A.}~\bibnamefont {Meerschaut}}\ and\ \bibinfo {author} {\bibfnamefont {J.}~\bibnamefont {Rouxel}},\ }\bibinfo {title} {Pseudo-one-dimensional mx3 and mx4 transition metal chalcogenides},\ in\ \href {https://doi.org/10.1007/978-94-009-4528-9_6} {\emph {\bibinfo {booktitle} {Crystal Chemistry and Properties of Materials with Quasi-One-Dimensional Structures: A Chemical and Physical Synthetic Approach}}},\ \bibinfo {editor} {edited by\ \bibinfo {editor} {\bibfnamefont {J.}~\bibnamefont {Rouxel}}}\ (\bibinfo  {publisher} {Springer Netherlands},\ \bibinfo {address} {Dordrecht},\ \bibinfo {year} {1986})\ pp.\ \bibinfo {pages} {205--279}\BibitemShut {NoStop}%
\bibitem [{\citenamefont {Lieth}\ and\ \citenamefont {Terhell}(1977)}]{IntroLibro2}%
  \BibitemOpen
  \bibfield  {author} {\bibinfo {author} {\bibfnamefont {R.~M.~A.}\ \bibnamefont {Lieth}}\ and\ \bibinfo {author} {\bibfnamefont {J.~C. J.~M.}\ \bibnamefont {Terhell}},\ }\bibinfo {title} {Transition metal dichalcogenides},\ in\ \href {https://doi.org/10.1007/978-94-017-2750-1_4} {\emph {\bibinfo {booktitle} {Preparation and Crystal Growth of Materials with Layered Structures}}},\ \bibinfo {editor} {edited by\ \bibinfo {editor} {\bibfnamefont {R.~M.~A.}\ \bibnamefont {Lieth}}}\ (\bibinfo  {publisher} {Springer Netherlands},\ \bibinfo {address} {Dordrecht},\ \bibinfo {year} {1977})\ pp.\ \bibinfo {pages} {141--223}\BibitemShut {NoStop}%
\bibitem [{\citenamefont {Castro~Neto}(2001)}]{CDWgeneral}%
  \BibitemOpen
  \bibfield  {author} {\bibinfo {author} {\bibfnamefont {A.~H.}\ \bibnamefont {Castro~Neto}},\ }\bibfield  {title} {\bibinfo {title} {Charge density wave, superconductivity, and anomalous metallic behavior in 2d transition metal dichalcogenides},\ }\href {https://doi.org/10.1103/PhysRevLett.86.4382} {\bibfield  {journal} {\bibinfo  {journal} {Phys. Rev. Lett.}\ }\textbf {\bibinfo {volume} {86}},\ \bibinfo {pages} {4382} (\bibinfo {year} {2001})}\BibitemShut {NoStop}%
\bibitem [{\citenamefont {Klemm}(1985)}]{Superconductivitygeneral}%
  \BibitemOpen
  \bibfield  {author} {\bibinfo {author} {\bibfnamefont {R.~A.}\ \bibnamefont {Klemm}},\ }\bibinfo {title} {Theory of the superconducting properties of quasi-one-dimensional materials},\ in\ \href {https://doi.org/10.1007/978-94-015-6923-1_5} {\emph {\bibinfo {booktitle} {Electronic Properties of Inorganic Quasi-One-Dimensional Compounds: Part I --- Theoretical}}},\ \bibinfo {editor} {edited by\ \bibinfo {editor} {\bibfnamefont {P.}~\bibnamefont {Monceau}}}\ (\bibinfo  {publisher} {Springer Netherlands},\ \bibinfo {address} {Dordrecht},\ \bibinfo {year} {1985})\ pp.\ \bibinfo {pages} {195--241}\BibitemShut {NoStop}%
\bibitem [{\citenamefont {Zhang}\ \emph {et~al.}(2021{\natexlab{a}})\citenamefont {Zhang}, \citenamefont {Ding}, \citenamefont {Chen}, \citenamefont {Guo}, \citenamefont {Pan}, \citenamefont {Li}, \citenamefont {Zhao}, \citenamefont {Liu},\ and\ \citenamefont {Xie}}]{Magnetismgeneral}%
  \BibitemOpen
  \bibfield  {author} {\bibinfo {author} {\bibfnamefont {Y.}~\bibnamefont {Zhang}}, \bibinfo {author} {\bibfnamefont {W.}~\bibnamefont {Ding}}, \bibinfo {author} {\bibfnamefont {Z.}~\bibnamefont {Chen}}, \bibinfo {author} {\bibfnamefont {J.}~\bibnamefont {Guo}}, \bibinfo {author} {\bibfnamefont {H.}~\bibnamefont {Pan}}, \bibinfo {author} {\bibfnamefont {X.}~\bibnamefont {Li}}, \bibinfo {author} {\bibfnamefont {Z.}~\bibnamefont {Zhao}}, \bibinfo {author} {\bibfnamefont {Y.}~\bibnamefont {Liu}},\ and\ \bibinfo {author} {\bibfnamefont {W.}~\bibnamefont {Xie}},\ }\bibfield  {title} {\bibinfo {title} {Layer-dependent magnetism in two-dimensional transition-metal chalcogenides mntn + 1 (m = v, cr, and mn; t = s, se, and te; and n = 2, 3, and 4)},\ }\href {https://doi.org/10.1021/acs.jpcc.0c11449} {\bibfield  {journal} {\bibinfo  {journal} {The Journal of Physical Chemistry C}\ }\textbf {\bibinfo {volume} {125}},\ \bibinfo {pages} {8398} (\bibinfo {year} {2021}{\natexlab{a}})}\BibitemShut {NoStop}%
\bibitem [{\citenamefont {AlBuhairan}\ and\ \citenamefont {Vogl}(2023)}]{PhysRevB.108.155106}%
  \BibitemOpen
  \bibfield  {author} {\bibinfo {author} {\bibfnamefont {H.}~\bibnamefont {AlBuhairan}}\ and\ \bibinfo {author} {\bibfnamefont {M.}~\bibnamefont {Vogl}},\ }\bibfield  {title} {\bibinfo {title} {Band structure and band topology in twisted homotrilayer transition metal dichalcogenides},\ }\href {https://doi.org/10.1103/PhysRevB.108.155106} {\bibfield  {journal} {\bibinfo  {journal} {Phys. Rev. B}\ }\textbf {\bibinfo {volume} {108}},\ \bibinfo {pages} {155106} (\bibinfo {year} {2023})}\BibitemShut {NoStop}%
\bibitem [{\citenamefont {Zhang}\ \emph {et~al.}(2021{\natexlab{b}})\citenamefont {Zhang}, \citenamefont {Zhang}, \citenamefont {Qiu}, \citenamefont {Zhao},\ and\ \citenamefont {Liu}}]{doi.org/10.1002/inf2.12156}%
  \BibitemOpen
  \bibfield  {author} {\bibinfo {author} {\bibfnamefont {W.}~\bibnamefont {Zhang}}, \bibinfo {author} {\bibfnamefont {Y.}~\bibnamefont {Zhang}}, \bibinfo {author} {\bibfnamefont {J.}~\bibnamefont {Qiu}}, \bibinfo {author} {\bibfnamefont {Z.}~\bibnamefont {Zhao}},\ and\ \bibinfo {author} {\bibfnamefont {N.}~\bibnamefont {Liu}},\ }\bibfield  {title} {\bibinfo {title} {Topological structures of transition metal dichalcogenides: A review on fabrication, effects, applications, and potential},\ }\href {https://doi.org/https://doi.org/10.1002/inf2.12156} {\bibfield  {journal} {\bibinfo  {journal} {InfoMat}\ }\textbf {\bibinfo {volume} {3}},\ \bibinfo {pages} {133} (\bibinfo {year} {2021}{\natexlab{b}})}\BibitemShut {NoStop}%
\bibitem [{\citenamefont {Zhang}\ \emph {et~al.}(2020)\citenamefont {Zhang}, \citenamefont {Gu}, \citenamefont {Sun}, \citenamefont {Luo}, \citenamefont {Liu}, \citenamefont {Chen}, \citenamefont {Shao}, \citenamefont {Zhang}, \citenamefont {Li}, \citenamefont {Sun}, \citenamefont {Li}, \citenamefont {Li}, \citenamefont {Xue}, \citenamefont {Ge}, \citenamefont {Xing}, \citenamefont {Comin}, \citenamefont {Zhu}, \citenamefont {Gao}, \citenamefont {Yan}, \citenamefont {Feng}, \citenamefont {Pan},\ and\ \citenamefont {Wang}}]{PhysRevB.102.035125}%
  \BibitemOpen
  \bibfield  {author} {\bibinfo {author} {\bibfnamefont {X.}~\bibnamefont {Zhang}}, \bibinfo {author} {\bibfnamefont {Q.}~\bibnamefont {Gu}}, \bibinfo {author} {\bibfnamefont {H.}~\bibnamefont {Sun}}, \bibinfo {author} {\bibfnamefont {T.}~\bibnamefont {Luo}}, \bibinfo {author} {\bibfnamefont {Y.}~\bibnamefont {Liu}}, \bibinfo {author} {\bibfnamefont {Y.}~\bibnamefont {Chen}}, \bibinfo {author} {\bibfnamefont {Z.}~\bibnamefont {Shao}}, \bibinfo {author} {\bibfnamefont {Z.}~\bibnamefont {Zhang}}, \bibinfo {author} {\bibfnamefont {S.}~\bibnamefont {Li}}, \bibinfo {author} {\bibfnamefont {Y.}~\bibnamefont {Sun}}, \bibinfo {author} {\bibfnamefont {Y.}~\bibnamefont {Li}}, \bibinfo {author} {\bibfnamefont {X.}~\bibnamefont {Li}}, \bibinfo {author} {\bibfnamefont {S.}~\bibnamefont {Xue}}, \bibinfo {author} {\bibfnamefont {J.}~\bibnamefont {Ge}}, \bibinfo {author} {\bibfnamefont {Y.}~\bibnamefont {Xing}}, \bibinfo {author} {\bibfnamefont {R.}~\bibnamefont {Comin}}, \bibinfo {author} {\bibfnamefont {Z.}~\bibnamefont
  {Zhu}}, \bibinfo {author} {\bibfnamefont {P.}~\bibnamefont {Gao}}, \bibinfo {author} {\bibfnamefont {B.}~\bibnamefont {Yan}}, \bibinfo {author} {\bibfnamefont {J.}~\bibnamefont {Feng}}, \bibinfo {author} {\bibfnamefont {M.}~\bibnamefont {Pan}},\ and\ \bibinfo {author} {\bibfnamefont {J.}~\bibnamefont {Wang}},\ }\bibfield  {title} {\bibinfo {title} {Eightfold fermionic excitation in a charge density wave compound},\ }\href {https://doi.org/10.1103/PhysRevB.102.035125} {\bibfield  {journal} {\bibinfo  {journal} {Phys. Rev. B}\ }\textbf {\bibinfo {volume} {102}},\ \bibinfo {pages} {035125} (\bibinfo {year} {2020})}\BibitemShut {NoStop}%
\bibitem [{\citenamefont {Daschner}\ and\ \citenamefont {Grosche}(2025)}]{daschner2025electronholetunnellingprobedhaas}%
  \BibitemOpen
  \bibfield  {author} {\bibinfo {author} {\bibfnamefont {M.}~\bibnamefont {Daschner}}\ and\ \bibinfo {author} {\bibfnamefont {F.~M.}\ \bibnamefont {Grosche}},\ }\href {https://arxiv.org/abs/2507.02579} {\bibinfo {title} {Electron-hole tunnelling probed in de haas - van alphen oscillations in the (double) dirac semimetal nbte$_4$}} (\bibinfo {year} {2025}),\ \Eprint {https://arxiv.org/abs/2507.02579} {arXiv:2507.02579 [cond-mat.mtrl-sci]} \BibitemShut {NoStop}%
\bibitem [{\citenamefont {Rezende-Gon\ifmmode~\mbox{\c{c}}\else \c{c}\fi{}alves}\ \emph {et~al.}(2024)\citenamefont {Rezende-Gon\ifmmode~\mbox{\c{c}}\else \c{c}\fi{}alves}, \citenamefont {Thees}, \citenamefont {Rojas-Castillo}, \citenamefont {Silvera-Vega}, \citenamefont {Bouwmeester}, \citenamefont {David}, \citenamefont {Antezak}, \citenamefont {Thakur}, \citenamefont {Fortuna}, \citenamefont {Le~F\`evre}, \citenamefont {Rosmus}, \citenamefont {Olszowska}, \citenamefont {Sobol}, \citenamefont {Magalh\~aes Paniago}, \citenamefont {Garcia-Castro}, \citenamefont {Giraldo-Gallo}, \citenamefont {Frantzeskakis},\ and\ \citenamefont {Santander-Syro}}]{PhysRevB.110.125151}%
  \BibitemOpen
  \bibfield  {author} {\bibinfo {author} {\bibfnamefont {P.}~\bibnamefont {Rezende-Gon\ifmmode~\mbox{\c{c}}\else \c{c}\fi{}alves}}, \bibinfo {author} {\bibfnamefont {M.}~\bibnamefont {Thees}}, \bibinfo {author} {\bibfnamefont {J.}~\bibnamefont {Rojas-Castillo}}, \bibinfo {author} {\bibfnamefont {D.}~\bibnamefont {Silvera-Vega}}, \bibinfo {author} {\bibfnamefont {R.~L.}\ \bibnamefont {Bouwmeester}}, \bibinfo {author} {\bibfnamefont {E.}~\bibnamefont {David}}, \bibinfo {author} {\bibfnamefont {A.}~\bibnamefont {Antezak}}, \bibinfo {author} {\bibfnamefont {A.~J.}\ \bibnamefont {Thakur}}, \bibinfo {author} {\bibfnamefont {F.}~\bibnamefont {Fortuna}}, \bibinfo {author} {\bibfnamefont {P.}~\bibnamefont {Le~F\`evre}}, \bibinfo {author} {\bibfnamefont {M.}~\bibnamefont {Rosmus}}, \bibinfo {author} {\bibfnamefont {N.}~\bibnamefont {Olszowska}}, \bibinfo {author} {\bibfnamefont {T.}~\bibnamefont {Sobol}}, \bibinfo {author} {\bibfnamefont {R.}~\bibnamefont {Magalh\~aes Paniago}}, \bibinfo {author} {\bibfnamefont
  {A.~C.}\ \bibnamefont {Garcia-Castro}}, \bibinfo {author} {\bibfnamefont {P.}~\bibnamefont {Giraldo-Gallo}}, \bibinfo {author} {\bibfnamefont {E.}~\bibnamefont {Frantzeskakis}},\ and\ \bibinfo {author} {\bibfnamefont {A.~F.}\ \bibnamefont {Santander-Syro}},\ }\bibfield  {title} {\bibinfo {title} {Experimental observation of metallic states with different dimensionality in a quasi-one-dimensional charge density wave compound},\ }\href {https://doi.org/10.1103/PhysRevB.110.125151} {\bibfield  {journal} {\bibinfo  {journal} {Phys. Rev. B}\ }\textbf {\bibinfo {volume} {110}},\ \bibinfo {pages} {125151} (\bibinfo {year} {2024})}\BibitemShut {NoStop}%
\bibitem [{\citenamefont {Xu}\ \emph {et~al.}(2023)\citenamefont {Xu}, \citenamefont {Du}, \citenamefont {Zhou}, \citenamefont {Gu}, \citenamefont {Zhang}, \citenamefont {Li}, \citenamefont {Zhao}, \citenamefont {Zheng}, \citenamefont {Arita}, \citenamefont {Shimada}, \citenamefont {Kim}, \citenamefont {Cacho}, \citenamefont {Guo}, \citenamefont {Liu}, \citenamefont {Chen},\ and\ \citenamefont {Yang}}]{Xu2023}%
  \BibitemOpen
  \bibfield  {author} {\bibinfo {author} {\bibfnamefont {R.~Z.}\ \bibnamefont {Xu}}, \bibinfo {author} {\bibfnamefont {X.}~\bibnamefont {Du}}, \bibinfo {author} {\bibfnamefont {J.~S.}\ \bibnamefont {Zhou}}, \bibinfo {author} {\bibfnamefont {X.}~\bibnamefont {Gu}}, \bibinfo {author} {\bibfnamefont {Q.~Q.}\ \bibnamefont {Zhang}}, \bibinfo {author} {\bibfnamefont {Y.~D.}\ \bibnamefont {Li}}, \bibinfo {author} {\bibfnamefont {W.~X.}\ \bibnamefont {Zhao}}, \bibinfo {author} {\bibfnamefont {F.~W.}\ \bibnamefont {Zheng}}, \bibinfo {author} {\bibfnamefont {M.}~\bibnamefont {Arita}}, \bibinfo {author} {\bibfnamefont {K.}~\bibnamefont {Shimada}}, \bibinfo {author} {\bibfnamefont {T.~K.}\ \bibnamefont {Kim}}, \bibinfo {author} {\bibfnamefont {C.}~\bibnamefont {Cacho}}, \bibinfo {author} {\bibfnamefont {Y.~F.}\ \bibnamefont {Guo}}, \bibinfo {author} {\bibfnamefont {Z.~K.}\ \bibnamefont {Liu}}, \bibinfo {author} {\bibfnamefont {Y.~L.}\ \bibnamefont {Chen}},\ and\ \bibinfo {author} {\bibfnamefont {L.~X.}\ \bibnamefont
  {Yang}},\ }\bibfield  {title} {\bibinfo {title} {Orbital-selective charge-density wave in tate4},\ }\href {https://doi.org/10.1038/s41535-023-00573-8} {\bibfield  {journal} {\bibinfo  {journal} {npj Quantum Materials}\ }\textbf {\bibinfo {volume} {8}},\ \bibinfo {pages} {44} (\bibinfo {year} {2023})}\BibitemShut {NoStop}%
\bibitem [{\citenamefont {Rong}\ \emph {et~al.}(2024)\citenamefont {Rong}, \citenamefont {Gu}, \citenamefont {Yang}, \citenamefont {Lin}, \citenamefont {Zhang}, \citenamefont {Wang}, \citenamefont {Guo}, \citenamefont {Hao}, \citenamefont {Cai}, \citenamefont {Zhang}, \citenamefont {Jiang}, \citenamefont {Yang}, \citenamefont {Jiang}, \citenamefont {Liu}, \citenamefont {Ye}, \citenamefont {Guo}, \citenamefont {Wang}, \citenamefont {Lin}, \citenamefont {Lu}, \citenamefont {Liu}, \citenamefont {Feng},\ and\ \citenamefont {Chen}}]{PhysRevLett.133.116403}%
  \BibitemOpen
  \bibfield  {author} {\bibinfo {author} {\bibfnamefont {H.}~\bibnamefont {Rong}}, \bibinfo {author} {\bibfnamefont {Q.}~\bibnamefont {Gu}}, \bibinfo {author} {\bibfnamefont {X.}~\bibnamefont {Yang}}, \bibinfo {author} {\bibfnamefont {Y.}~\bibnamefont {Lin}}, \bibinfo {author} {\bibfnamefont {C.}~\bibnamefont {Zhang}}, \bibinfo {author} {\bibfnamefont {Y.}~\bibnamefont {Wang}}, \bibinfo {author} {\bibfnamefont {R.}~\bibnamefont {Guo}}, \bibinfo {author} {\bibfnamefont {Z.}~\bibnamefont {Hao}}, \bibinfo {author} {\bibfnamefont {Y.}~\bibnamefont {Cai}}, \bibinfo {author} {\bibfnamefont {F.}~\bibnamefont {Zhang}}, \bibinfo {author} {\bibfnamefont {Z.}~\bibnamefont {Jiang}}, \bibinfo {author} {\bibfnamefont {Y.}~\bibnamefont {Yang}}, \bibinfo {author} {\bibfnamefont {Q.}~\bibnamefont {Jiang}}, \bibinfo {author} {\bibfnamefont {Z.}~\bibnamefont {Liu}}, \bibinfo {author} {\bibfnamefont {M.}~\bibnamefont {Ye}}, \bibinfo {author} {\bibfnamefont {S.}~\bibnamefont {Guo}}, \bibinfo {author} {\bibfnamefont
  {L.}~\bibnamefont {Wang}}, \bibinfo {author} {\bibfnamefont {J.}~\bibnamefont {Lin}}, \bibinfo {author} {\bibfnamefont {L.}~\bibnamefont {Lu}}, \bibinfo {author} {\bibfnamefont {G.}~\bibnamefont {Liu}}, \bibinfo {author} {\bibfnamefont {J.}~\bibnamefont {Feng}},\ and\ \bibinfo {author} {\bibfnamefont {C.}~\bibnamefont {Chen}},\ }\bibfield  {title} {\bibinfo {title} {Dominant charge density order in ${\mathrm{tate}}_{4}$},\ }\href {https://doi.org/10.1103/PhysRevLett.133.116403} {\bibfield  {journal} {\bibinfo  {journal} {Phys. Rev. Lett.}\ }\textbf {\bibinfo {volume} {133}},\ \bibinfo {pages} {116403} (\bibinfo {year} {2024})}\BibitemShut {NoStop}%
\bibitem [{\citenamefont {Bennett}\ and\ \citenamefont {Boswell}(1999)}]{CDWmodulations}%
  \BibitemOpen
  \bibfield  {author} {\bibinfo {author} {\bibfnamefont {J.~C.}\ \bibnamefont {Bennett}}\ and\ \bibinfo {author} {\bibfnamefont {F.~W.}\ \bibnamefont {Boswell}},\ }\bibinfo {title} {Charge density wave phase transitions and microstructures in the tate4 --- nbte4 system},\ in\ \href {https://doi.org/10.1007/978-94-011-4603-6_3} {\emph {\bibinfo {booktitle} {Advances in the Crystallographic and Microstructural Analysis of Charge Density Wave Modulated Crystals}}},\ \bibinfo {editor} {edited by\ \bibinfo {editor} {\bibfnamefont {F.~W.}\ \bibnamefont {Boswell}}\ and\ \bibinfo {editor} {\bibfnamefont {J.~C.}\ \bibnamefont {Bennett}}}\ (\bibinfo  {publisher} {Springer Netherlands},\ \bibinfo {address} {Dordrecht},\ \bibinfo {year} {1999})\ pp.\ \bibinfo {pages} {69--120}\BibitemShut {NoStop}%
\bibitem [{\citenamefont {Yang}\ \emph {et~al.}(2018)\citenamefont {Yang}, \citenamefont {Zhou}, \citenamefont {Wang}, \citenamefont {Bai}, \citenamefont {Chen}, \citenamefont {An}, \citenamefont {Zhou}, \citenamefont {Chen}, \citenamefont {Li}, \citenamefont {Wang}, \citenamefont {Chen}, \citenamefont {Cao}, \citenamefont {Li}, \citenamefont {Zhou}, \citenamefont {Yang},\ and\ \citenamefont {Xu}}]{pressuresuperconductivityNbTe4}%
  \BibitemOpen
  \bibfield  {author} {\bibinfo {author} {\bibfnamefont {X.}~\bibnamefont {Yang}}, \bibinfo {author} {\bibfnamefont {Y.}~\bibnamefont {Zhou}}, \bibinfo {author} {\bibfnamefont {M.}~\bibnamefont {Wang}}, \bibinfo {author} {\bibfnamefont {H.}~\bibnamefont {Bai}}, \bibinfo {author} {\bibfnamefont {X.}~\bibnamefont {Chen}}, \bibinfo {author} {\bibfnamefont {C.}~\bibnamefont {An}}, \bibinfo {author} {\bibfnamefont {Y.}~\bibnamefont {Zhou}}, \bibinfo {author} {\bibfnamefont {Q.}~\bibnamefont {Chen}}, \bibinfo {author} {\bibfnamefont {Y.}~\bibnamefont {Li}}, \bibinfo {author} {\bibfnamefont {Z.}~\bibnamefont {Wang}}, \bibinfo {author} {\bibfnamefont {J.}~\bibnamefont {Chen}}, \bibinfo {author} {\bibfnamefont {C.}~\bibnamefont {Cao}}, \bibinfo {author} {\bibfnamefont {Y.}~\bibnamefont {Li}}, \bibinfo {author} {\bibfnamefont {Y.}~\bibnamefont {Zhou}}, \bibinfo {author} {\bibfnamefont {Z.}~\bibnamefont {Yang}},\ and\ \bibinfo {author} {\bibfnamefont {Z.-A.}\ \bibnamefont {Xu}},\ }\bibfield  {title} {\bibinfo {title}
  {Pressure induced superconductivity bordering a charge-density-wave state in nbte4 with strong spin-orbit coupling},\ }\href {https://doi.org/10.1038/s41598-018-24572-z} {\bibfield  {journal} {\bibinfo  {journal} {Scientific Reports}\ }\textbf {\bibinfo {volume} {8}},\ \bibinfo {pages} {6298} (\bibinfo {year} {2018})}\BibitemShut {NoStop}%
\bibitem [{\citenamefont {Yuan}\ \emph {et~al.}(2020)\citenamefont {Yuan}, \citenamefont {Wang}, \citenamefont {Zhou}, \citenamefont {Chen}, \citenamefont {Gu}, \citenamefont {An}, \citenamefont {Zhou}, \citenamefont {Zhang}, \citenamefont {Chen}, \citenamefont {Zhang},\ and\ \citenamefont {Yang}}]{pressuresuperconductivityTaTe4}%
  \BibitemOpen
  \bibfield  {author} {\bibinfo {author} {\bibfnamefont {Y.}~\bibnamefont {Yuan}}, \bibinfo {author} {\bibfnamefont {W.}~\bibnamefont {Wang}}, \bibinfo {author} {\bibfnamefont {Y.}~\bibnamefont {Zhou}}, \bibinfo {author} {\bibfnamefont {X.}~\bibnamefont {Chen}}, \bibinfo {author} {\bibfnamefont {C.}~\bibnamefont {Gu}}, \bibinfo {author} {\bibfnamefont {C.}~\bibnamefont {An}}, \bibinfo {author} {\bibfnamefont {Y.}~\bibnamefont {Zhou}}, \bibinfo {author} {\bibfnamefont {B.}~\bibnamefont {Zhang}}, \bibinfo {author} {\bibfnamefont {C.}~\bibnamefont {Chen}}, \bibinfo {author} {\bibfnamefont {R.}~\bibnamefont {Zhang}},\ and\ \bibinfo {author} {\bibfnamefont {Z.}~\bibnamefont {Yang}},\ }\bibfield  {title} {\bibinfo {title} {Pressure-induced superconductivity in topological semimetal candidate tate4},\ }\href {https://doi.org/https://doi.org/10.1002/aelm.201901260} {\bibfield  {journal} {\bibinfo  {journal} {Advanced Electronic Materials}\ }\textbf {\bibinfo {volume} {6}},\ \bibinfo {pages} {1901260} (\bibinfo {year}
  {2020})}\BibitemShut {NoStop}%
\bibitem [{\citenamefont {Gao}\ \emph {et~al.}(2017)\citenamefont {Gao}, \citenamefont {Xu}, \citenamefont {Qiu}, \citenamefont {Tian}, \citenamefont {Yuan},\ and\ \citenamefont {Wang}}]{10.1063/1.5005907}%
  \BibitemOpen
  \bibfield  {author} {\bibinfo {author} {\bibfnamefont {Y.}~\bibnamefont {Gao}}, \bibinfo {author} {\bibfnamefont {L.}~\bibnamefont {Xu}}, \bibinfo {author} {\bibfnamefont {Y.}~\bibnamefont {Qiu}}, \bibinfo {author} {\bibfnamefont {Z.}~\bibnamefont {Tian}}, \bibinfo {author} {\bibfnamefont {S.}~\bibnamefont {Yuan}},\ and\ \bibinfo {author} {\bibfnamefont {J.}~\bibnamefont {Wang}},\ }\bibfield  {title} {\bibinfo {title} {{Anisotropic large magnetoresistance in TaTe4 single crystals}},\ }\href {https://doi.org/10.1063/1.5005907} {\bibfield  {journal} {\bibinfo  {journal} {Journal of Applied Physics}\ }\textbf {\bibinfo {volume} {122}},\ \bibinfo {pages} {135101} (\bibinfo {year} {2017})}\BibitemShut {NoStop}%
\bibitem [{\citenamefont {Tadaki}\ \emph {et~al.}(1990)\citenamefont {Tadaki}, \citenamefont {Hino}, \citenamefont {Sambongi}, \citenamefont {Nomura},\ and\ \citenamefont {Lévy}}]{TADAKI1990227}%
  \BibitemOpen
  \bibfield  {author} {\bibinfo {author} {\bibfnamefont {S.}~\bibnamefont {Tadaki}}, \bibinfo {author} {\bibfnamefont {N.}~\bibnamefont {Hino}}, \bibinfo {author} {\bibfnamefont {T.}~\bibnamefont {Sambongi}}, \bibinfo {author} {\bibfnamefont {K.}~\bibnamefont {Nomura}},\ and\ \bibinfo {author} {\bibfnamefont {F.}~\bibnamefont {Lévy}},\ }\bibfield  {title} {\bibinfo {title} {Electrical properties of nbte4 and tate4},\ }\href {https://doi.org/https://doi.org/10.1016/0379-6779(90)90107-V} {\bibfield  {journal} {\bibinfo  {journal} {Synthetic Metals}\ }\textbf {\bibinfo {volume} {38}},\ \bibinfo {pages} {227} (\bibinfo {year} {1990})}\BibitemShut {NoStop}%
\bibitem [{\citenamefont {Boswell}\ \emph {et~al.}(1983{\natexlab{a}})\citenamefont {Boswell}, \citenamefont {Prodan},\ and\ \citenamefont {Brandon}}]{Boswell_1983}%
  \BibitemOpen
  \bibfield  {author} {\bibinfo {author} {\bibfnamefont {F.~W.}\ \bibnamefont {Boswell}}, \bibinfo {author} {\bibfnamefont {A.}~\bibnamefont {Prodan}},\ and\ \bibinfo {author} {\bibfnamefont {J.~K.}\ \bibnamefont {Brandon}},\ }\bibfield  {title} {\bibinfo {title} {Charge-density waves in the quasi-one-dimensional compounds nbte4 and tate4},\ }\href {https://doi.org/10.1088/0022-3719/16/6/012} {\bibfield  {journal} {\bibinfo  {journal} {Journal of Physics C: Solid State Physics}\ }\textbf {\bibinfo {volume} {16}},\ \bibinfo {pages} {1067} (\bibinfo {year} {1983}{\natexlab{a}})}\BibitemShut {NoStop}%
\bibitem [{\citenamefont {Boswell}\ and\ \citenamefont {Prodan}(1984)}]{BOSWELL198493}%
  \BibitemOpen
  \bibfield  {author} {\bibinfo {author} {\bibfnamefont {F.}~\bibnamefont {Boswell}}\ and\ \bibinfo {author} {\bibfnamefont {A.}~\bibnamefont {Prodan}},\ }\bibfield  {title} {\bibinfo {title} {Charge-density-waves of variable incommensurability in the system tate4 -- nbte4},\ }\href {https://doi.org/https://doi.org/10.1016/0025-5408(84)90014-X} {\bibfield  {journal} {\bibinfo  {journal} {Materials Research Bulletin}\ }\textbf {\bibinfo {volume} {19}},\ \bibinfo {pages} {93} (\bibinfo {year} {1984})}\BibitemShut {NoStop}%
\bibitem [{\citenamefont {Mahy}\ \emph {et~al.}(1983{\natexlab{a}})\citenamefont {Mahy}, \citenamefont {Van~Landuyt},\ and\ \citenamefont {Amelinckx}}]{https://doi.org/10.1002/pssa.2210770151}%
  \BibitemOpen
  \bibfield  {author} {\bibinfo {author} {\bibfnamefont {J.}~\bibnamefont {Mahy}}, \bibinfo {author} {\bibfnamefont {J.}~\bibnamefont {Van~Landuyt}},\ and\ \bibinfo {author} {\bibfnamefont {S.}~\bibnamefont {Amelinckx}},\ }\bibfield  {title} {\bibinfo {title} {Electron diffraction evidence for superstructures in tate4 and nbte4},\ }\href {https://doi.org/https://doi.org/10.1002/pssa.2210770151} {\bibfield  {journal} {\bibinfo  {journal} {physica status solidi (a)}\ }\textbf {\bibinfo {volume} {77}},\ \bibinfo {pages} {K1} (\bibinfo {year} {1983}{\natexlab{a}})}\BibitemShut {NoStop}%
\bibitem [{\citenamefont {Mahy}\ \emph {et~al.}(1983{\natexlab{b}})\citenamefont {Mahy}, \citenamefont {Wiegers}, \citenamefont {Van~Landuyt},\ and\ \citenamefont {Amelinckx}}]{mahy1983evidence}%
  \BibitemOpen
  \bibfield  {author} {\bibinfo {author} {\bibfnamefont {J.}~\bibnamefont {Mahy}}, \bibinfo {author} {\bibfnamefont {G.~A.}\ \bibnamefont {Wiegers}}, \bibinfo {author} {\bibfnamefont {J.}~\bibnamefont {Van~Landuyt}},\ and\ \bibinfo {author} {\bibfnamefont {S.}~\bibnamefont {Amelinckx}},\ }\bibfield  {title} {\bibinfo {title} {Evidence for deformation modulated structures in nbte$_4$ and tate$_4$},\ }\href {https://doi.org/10.1557/PROC-21-181} {\bibfield  {journal} {\bibinfo  {journal} {MRS Online Proceedings Library}\ }\textbf {\bibinfo {volume} {21}},\ \bibinfo {pages} {181} (\bibinfo {year} {1983}{\natexlab{b}})}\BibitemShut {NoStop}%
\bibitem [{\citenamefont {Boswell}\ \emph {et~al.}(1983{\natexlab{b}})\citenamefont {Boswell}, \citenamefont {Prodan},\ and\ \citenamefont {Brandon}}]{FWBoswell_1983}%
  \BibitemOpen
  \bibfield  {author} {\bibinfo {author} {\bibfnamefont {F.~W.}\ \bibnamefont {Boswell}}, \bibinfo {author} {\bibfnamefont {A.}~\bibnamefont {Prodan}},\ and\ \bibinfo {author} {\bibfnamefont {J.~K.}\ \bibnamefont {Brandon}},\ }\bibfield  {title} {\bibinfo {title} {Charge-density waves in the quasi-one-dimensional compounds nbte4 and tate4},\ }\href {https://doi.org/10.1088/0022-3719/16/6/012} {\bibfield  {journal} {\bibinfo  {journal} {Journal of Physics C: Solid State Physics}\ }\textbf {\bibinfo {volume} {16}},\ \bibinfo {pages} {1067} (\bibinfo {year} {1983}{\natexlab{b}})}\BibitemShut {NoStop}%
\bibitem [{\citenamefont {Boswell}\ and\ \citenamefont {Prodan}(1986)}]{PhysRevB.34.2979}%
  \BibitemOpen
  \bibfield  {author} {\bibinfo {author} {\bibfnamefont {F.~W.}\ \bibnamefont {Boswell}}\ and\ \bibinfo {author} {\bibfnamefont {A.}~\bibnamefont {Prodan}},\ }\bibfield  {title} {\bibinfo {title} {Structural changes in the discommensurate distortion waves of nb${\mathrm{te}}_{4}$ on cooling},\ }\href {https://doi.org/10.1103/PhysRevB.34.2979} {\bibfield  {journal} {\bibinfo  {journal} {Phys. Rev. B}\ }\textbf {\bibinfo {volume} {34}},\ \bibinfo {pages} {2979} (\bibinfo {year} {1986})}\BibitemShut {NoStop}%
\bibitem [{\citenamefont {van Smaalen}\ \emph {et~al.}(1986)\citenamefont {van Smaalen}, \citenamefont {Bronsema},\ and\ \citenamefont {Mahy}}]{vanSmaalen:a25451}%
  \BibitemOpen
  \bibfield  {author} {\bibinfo {author} {\bibfnamefont {S.}~\bibnamefont {van Smaalen}}, \bibinfo {author} {\bibfnamefont {K.~D.}\ \bibnamefont {Bronsema}},\ and\ \bibinfo {author} {\bibfnamefont {J.}~\bibnamefont {Mahy}},\ }\bibfield  {title} {\bibinfo {title} {{The determination of the incommensurately modulated structure of niobium tetratelluride}},\ }\href {https://doi.org/10.1107/S0108768186098609} {\bibfield  {journal} {\bibinfo  {journal} {Acta Crystallographica Section B}\ }\textbf {\bibinfo {volume} {42}},\ \bibinfo {pages} {43} (\bibinfo {year} {1986})}\BibitemShut {NoStop}%
\bibitem [{\citenamefont {Böhm}(1987)}]{Böhm+1987+113+122}%
  \BibitemOpen
  \bibfield  {author} {\bibinfo {author} {\bibfnamefont {H.}~\bibnamefont {Böhm}},\ }\bibfield  {title} {\bibinfo {title} {The high temperature modification of niobium tetratelluride nbte4},\ }\href {https://doi.org/doi:10.1524/zkri.1987.180.14.113} {\bibfield  {journal} {\bibinfo  {journal} {Zeitschrift für Kristallographie - Crystalline Materials}\ }\textbf {\bibinfo {volume} {180}},\ \bibinfo {pages} {113} (\bibinfo {year} {1987})}\BibitemShut {NoStop}%
\bibitem [{\citenamefont {{Galvis}}\ \emph {et~al.}(2023)\citenamefont {{Galvis}}, \citenamefont {{Fang}}, \citenamefont {{Jim{\'e}nez-Guerrero}}, \citenamefont {{Rojas-Castillo}}, \citenamefont {{Casas}}, \citenamefont {{Herrera}}, \citenamefont {{Garcia-Castro}}, \citenamefont {{Bousquet}}, \citenamefont {{Fisher}}, \citenamefont {{Kapitulnik}},\ and\ \citenamefont {{Giraldo-Gallo}}}]{2023PhRvB.107d5120G}%
  \BibitemOpen
  \bibfield  {author} {\bibinfo {author} {\bibfnamefont {J.~A.}\ \bibnamefont {{Galvis}}}, \bibinfo {author} {\bibfnamefont {A.}~\bibnamefont {{Fang}}}, \bibinfo {author} {\bibfnamefont {D.}~\bibnamefont {{Jim{\'e}nez-Guerrero}}}, \bibinfo {author} {\bibfnamefont {J.}~\bibnamefont {{Rojas-Castillo}}}, \bibinfo {author} {\bibfnamefont {J.}~\bibnamefont {{Casas}}}, \bibinfo {author} {\bibfnamefont {O.}~\bibnamefont {{Herrera}}}, \bibinfo {author} {\bibfnamefont {A.~C.}\ \bibnamefont {{Garcia-Castro}}}, \bibinfo {author} {\bibfnamefont {E.}~\bibnamefont {{Bousquet}}}, \bibinfo {author} {\bibfnamefont {I.~R.}\ \bibnamefont {{Fisher}}}, \bibinfo {author} {\bibfnamefont {A.}~\bibnamefont {{Kapitulnik}}},\ and\ \bibinfo {author} {\bibfnamefont {P.}~\bibnamefont {{Giraldo-Gallo}}},\ }\bibfield  {title} {\bibinfo {title} {{Nanoscale phase-slip domain walls in the charge density wave state of the Weyl semimetal candidate NbTe$_{4}$}},\ }\href {https://doi.org/10.1103/PhysRevB.107.045120} {\bibfield  {journal} {\bibinfo
   {journal} {\prb}\ }\textbf {\bibinfo {volume} {107}},\ \bibinfo {eid} {045120} (\bibinfo {year} {2023})}\BibitemShut {NoStop}%
\bibitem [{\citenamefont {Liu}\ \emph {et~al.}(2021)\citenamefont {Liu}, \citenamefont {Fu}, \citenamefont {Deng}, \citenamefont {Yuan},\ and\ \citenamefont {Wu}}]{10.1063/5.0053990}%
  \BibitemOpen
  \bibfield  {author} {\bibinfo {author} {\bibfnamefont {F.-H.}\ \bibnamefont {Liu}}, \bibinfo {author} {\bibfnamefont {W.}~\bibnamefont {Fu}}, \bibinfo {author} {\bibfnamefont {Y.-H.}\ \bibnamefont {Deng}}, \bibinfo {author} {\bibfnamefont {Z.-B.}\ \bibnamefont {Yuan}},\ and\ \bibinfo {author} {\bibfnamefont {L.-N.}\ \bibnamefont {Wu}},\ }\bibfield  {title} {\bibinfo {title} {First-principles study of the kohn anomaly in tate4},\ }\href {https://doi.org/10.1063/5.0053990} {\bibfield  {journal} {\bibinfo  {journal} {Applied Physics Letters}\ }\textbf {\bibinfo {volume} {119}},\ \bibinfo {pages} {091901} (\bibinfo {year} {2021})}\BibitemShut {NoStop}%
\bibitem [{\citenamefont {Guster}\ \emph {et~al.}(2022)\citenamefont {Guster}, \citenamefont {Pruneda}, \citenamefont {Ordej\'on},\ and\ \citenamefont {Canadell}}]{PhysRevB.105.064107}%
  \BibitemOpen
  \bibfield  {author} {\bibinfo {author} {\bibfnamefont {B.}~\bibnamefont {Guster}}, \bibinfo {author} {\bibfnamefont {M.}~\bibnamefont {Pruneda}}, \bibinfo {author} {\bibfnamefont {P.}~\bibnamefont {Ordej\'on}},\ and\ \bibinfo {author} {\bibfnamefont {E.}~\bibnamefont {Canadell}},\ }\bibfield  {title} {\bibinfo {title} {Competition between ta-ta and te-te bonding leading to the commensurate charge density wave in $\mathrm{Ta}{\mathrm{te}}_{4}$},\ }\href {https://doi.org/10.1103/PhysRevB.105.064107} {\bibfield  {journal} {\bibinfo  {journal} {Phys. Rev. B}\ }\textbf {\bibinfo {volume} {105}},\ \bibinfo {pages} {064107} (\bibinfo {year} {2022})}\BibitemShut {NoStop}%
\bibitem [{\citenamefont {Silvera-Vega}\ \emph {et~al.}(2026)\citenamefont {Silvera-Vega}, \citenamefont {Rojas-Castillo}, \citenamefont {Herrera-Vasco}, \citenamefont {Chikara}, \citenamefont {Ramos-Rodríguez}, \citenamefont {Herrera}, \citenamefont {Santander-Syro}, \citenamefont {Galvis}, \citenamefont {Uribe}, \citenamefont {González-Hernández}, \citenamefont {García-Castro},\ and\ \citenamefont {Giraldo-Gallo}}]{Silvera-Vega2026}%
  \BibitemOpen
  \bibfield  {author} {\bibinfo {author} {\bibfnamefont {D.}~\bibnamefont {Silvera-Vega}}, \bibinfo {author} {\bibfnamefont {J.}~\bibnamefont {Rojas-Castillo}}, \bibinfo {author} {\bibfnamefont {E.}~\bibnamefont {Herrera-Vasco}}, \bibinfo {author} {\bibfnamefont {S.}~\bibnamefont {Chikara}}, \bibinfo {author} {\bibfnamefont {E.}~\bibnamefont {Ramos-Rodríguez}}, \bibinfo {author} {\bibfnamefont {W.~J.}\ \bibnamefont {Herrera}}, \bibinfo {author} {\bibfnamefont {A.~F.}\ \bibnamefont {Santander-Syro}}, \bibinfo {author} {\bibfnamefont {J.~A.}\ \bibnamefont {Galvis}}, \bibinfo {author} {\bibfnamefont {B.}~\bibnamefont {Uribe}}, \bibinfo {author} {\bibfnamefont {R.}~\bibnamefont {González-Hernández}}, \bibinfo {author} {\bibfnamefont {A.~C.}\ \bibnamefont {García-Castro}},\ and\ \bibinfo {author} {\bibfnamefont {P.}~\bibnamefont {Giraldo-Gallo}},\ }\bibfield  {title} {\bibinfo {title} {Fermi surface reconstruction and anisotropic linear magnetoresistance in the charge density wave topological semimetal
  tate4},\ }\bibfield  {journal} {\bibinfo  {journal} {Communications Physics}\ }\href {https://doi.org/10.1038/s42005-026-02544-4} {10.1038/s42005-026-02544-4} (\bibinfo {year} {2026})\BibitemShut {NoStop}%
\bibitem [{\citenamefont {Luo}\ \emph {et~al.}(2022)\citenamefont {Luo}, \citenamefont {Gao}, \citenamefont {Liu}, \citenamefont {Gu}, \citenamefont {Wu}, \citenamefont {Yi}, \citenamefont {Jia}, \citenamefont {Wu}, \citenamefont {Luo}, \citenamefont {Xu}, \citenamefont {Zhao}, \citenamefont {Wang}, \citenamefont {Mao}, \citenamefont {Liu}, \citenamefont {Zhu}, \citenamefont {Shi}, \citenamefont {Jiang}, \citenamefont {Hu}, \citenamefont {Xu},\ and\ \citenamefont {Zhou}}]{luo2022electronic}%
  \BibitemOpen
  \bibfield  {author} {\bibinfo {author} {\bibfnamefont {H.}~\bibnamefont {Luo}}, \bibinfo {author} {\bibfnamefont {Q.}~\bibnamefont {Gao}}, \bibinfo {author} {\bibfnamefont {H.}~\bibnamefont {Liu}}, \bibinfo {author} {\bibfnamefont {Y.}~\bibnamefont {Gu}}, \bibinfo {author} {\bibfnamefont {D.}~\bibnamefont {Wu}}, \bibinfo {author} {\bibfnamefont {C.}~\bibnamefont {Yi}}, \bibinfo {author} {\bibfnamefont {J.}~\bibnamefont {Jia}}, \bibinfo {author} {\bibfnamefont {S.}~\bibnamefont {Wu}}, \bibinfo {author} {\bibfnamefont {X.}~\bibnamefont {Luo}}, \bibinfo {author} {\bibfnamefont {Y.}~\bibnamefont {Xu}}, \bibinfo {author} {\bibfnamefont {L.}~\bibnamefont {Zhao}}, \bibinfo {author} {\bibfnamefont {Q.}~\bibnamefont {Wang}}, \bibinfo {author} {\bibfnamefont {H.}~\bibnamefont {Mao}}, \bibinfo {author} {\bibfnamefont {G.}~\bibnamefont {Liu}}, \bibinfo {author} {\bibfnamefont {Z.}~\bibnamefont {Zhu}}, \bibinfo {author} {\bibfnamefont {Y.}~\bibnamefont {Shi}}, \bibinfo {author} {\bibfnamefont {K.}~\bibnamefont {Jiang}},
  \bibinfo {author} {\bibfnamefont {J.}~\bibnamefont {Hu}}, \bibinfo {author} {\bibfnamefont {Z.}~\bibnamefont {Xu}},\ and\ \bibinfo {author} {\bibfnamefont {X.~J.}\ \bibnamefont {Zhou}},\ }\bibfield  {title} {\bibinfo {title} {Electronic nature of charge density wave and electron-phonon coupling in kagome superconductor kv$_3$sb$_5$},\ }\href {https://doi.org/10.1038/s41467-021-27946-6} {\bibfield  {journal} {\bibinfo  {journal} {Nature Communications}\ }\textbf {\bibinfo {volume} {13}},\ \bibinfo {pages} {273} (\bibinfo {year} {2022})}\BibitemShut {NoStop}%
\bibitem [{\citenamefont {Gutierrez-Amigo}\ \emph {et~al.}(2024)\citenamefont {Gutierrez-Amigo}, \citenamefont {Dangi{\'c}}, \citenamefont {Guo}, \citenamefont {Felser}, \citenamefont {Moll}, \citenamefont {Vergniory},\ and\ \citenamefont {Errea}}]{gutierrez2024phonon}%
  \BibitemOpen
  \bibfield  {author} {\bibinfo {author} {\bibfnamefont {M.}~\bibnamefont {Gutierrez-Amigo}}, \bibinfo {author} {\bibfnamefont {D.}~\bibnamefont {Dangi{\'c}}}, \bibinfo {author} {\bibfnamefont {C.}~\bibnamefont {Guo}}, \bibinfo {author} {\bibfnamefont {C.}~\bibnamefont {Felser}}, \bibinfo {author} {\bibfnamefont {P.~J.~W.}\ \bibnamefont {Moll}}, \bibinfo {author} {\bibfnamefont {M.~G.}\ \bibnamefont {Vergniory}},\ and\ \bibinfo {author} {\bibfnamefont {I.}~\bibnamefont {Errea}},\ }\bibfield  {title} {\bibinfo {title} {Phonon collapse and anharmonic melting of the 3d charge-density wave in kagome metals},\ }\href {https://doi.org/10.1038/s43246-024-00676-0} {\bibfield  {journal} {\bibinfo  {journal} {Communications Materials}\ }\textbf {\bibinfo {volume} {5}},\ \bibinfo {pages} {234} (\bibinfo {year} {2024})}\BibitemShut {NoStop}%
\bibitem [{\citenamefont {Hu}\ \emph {et~al.}(2024)\citenamefont {Hu}, \citenamefont {Ma}, \citenamefont {Li}, \citenamefont {Jiang}, \citenamefont {Gawryluk}, \citenamefont {Hu}, \citenamefont {Teyssier}, \citenamefont {Multian}, \citenamefont {Yin}, \citenamefont {Xu}, \citenamefont {Shin}, \citenamefont {Plokhikh}, \citenamefont {Han}, \citenamefont {Plumb}, \citenamefont {Liu}, \citenamefont {Yin}, \citenamefont {Guguchia}, \citenamefont {Zhao}, \citenamefont {Schnyder}, \citenamefont {Wu}, \citenamefont {Pomjakushina}, \citenamefont {Hasan}, \citenamefont {Wang},\ and\ \citenamefont {Shi}}]{hu2024phonon}%
  \BibitemOpen
  \bibfield  {author} {\bibinfo {author} {\bibfnamefont {Y.}~\bibnamefont {Hu}}, \bibinfo {author} {\bibfnamefont {J.}~\bibnamefont {Ma}}, \bibinfo {author} {\bibfnamefont {Y.}~\bibnamefont {Li}}, \bibinfo {author} {\bibfnamefont {Y.}~\bibnamefont {Jiang}}, \bibinfo {author} {\bibfnamefont {D.~J.}\ \bibnamefont {Gawryluk}}, \bibinfo {author} {\bibfnamefont {T.}~\bibnamefont {Hu}}, \bibinfo {author} {\bibfnamefont {J.}~\bibnamefont {Teyssier}}, \bibinfo {author} {\bibfnamefont {V.}~\bibnamefont {Multian}}, \bibinfo {author} {\bibfnamefont {Z.}~\bibnamefont {Yin}}, \bibinfo {author} {\bibfnamefont {S.}~\bibnamefont {Xu}}, \bibinfo {author} {\bibfnamefont {S.}~\bibnamefont {Shin}}, \bibinfo {author} {\bibfnamefont {I.}~\bibnamefont {Plokhikh}}, \bibinfo {author} {\bibfnamefont {X.}~\bibnamefont {Han}}, \bibinfo {author} {\bibfnamefont {N.~C.}\ \bibnamefont {Plumb}}, \bibinfo {author} {\bibfnamefont {Y.}~\bibnamefont {Liu}}, \bibinfo {author} {\bibfnamefont {J.-X.}\ \bibnamefont {Yin}}, \bibinfo {author}
  {\bibfnamefont {Z.}~\bibnamefont {Guguchia}}, \bibinfo {author} {\bibfnamefont {Y.}~\bibnamefont {Zhao}}, \bibinfo {author} {\bibfnamefont {A.~P.}\ \bibnamefont {Schnyder}}, \bibinfo {author} {\bibfnamefont {X.}~\bibnamefont {Wu}}, \bibinfo {author} {\bibfnamefont {E.}~\bibnamefont {Pomjakushina}}, \bibinfo {author} {\bibfnamefont {M.~Z.}\ \bibnamefont {Hasan}}, \bibinfo {author} {\bibfnamefont {N.}~\bibnamefont {Wang}},\ and\ \bibinfo {author} {\bibfnamefont {M.}~\bibnamefont {Shi}},\ }\bibfield  {title} {\bibinfo {title} {Phonon promoted charge density wave in topological kagome metal scv$_6$sn$_6$},\ }\href {https://doi.org/10.1038/s41467-024-45859-y} {\bibfield  {journal} {\bibinfo  {journal} {Nature Communications}\ }\textbf {\bibinfo {volume} {15}},\ \bibinfo {pages} {1658} (\bibinfo {year} {2024})}\BibitemShut {NoStop}%
\bibitem [{\citenamefont {Hohenberg}\ and\ \citenamefont {Kohn}(1964)}]{PhysRev.136.B864}%
  \BibitemOpen
  \bibfield  {author} {\bibinfo {author} {\bibfnamefont {P.}~\bibnamefont {Hohenberg}}\ and\ \bibinfo {author} {\bibfnamefont {W.}~\bibnamefont {Kohn}},\ }\bibfield  {title} {\bibinfo {title} {Inhomogeneous electron gas},\ }\href {https://doi.org/10.1103/PhysRev.136.B864} {\bibfield  {journal} {\bibinfo  {journal} {Phys. Rev.}\ }\textbf {\bibinfo {volume} {136}},\ \bibinfo {pages} {B864} (\bibinfo {year} {1964})}\BibitemShut {NoStop}%
\bibitem [{\citenamefont {Kohn}\ and\ \citenamefont {Sham}(1965)}]{PhysRev.140.A1133}%
  \BibitemOpen
  \bibfield  {author} {\bibinfo {author} {\bibfnamefont {W.}~\bibnamefont {Kohn}}\ and\ \bibinfo {author} {\bibfnamefont {L.~J.}\ \bibnamefont {Sham}},\ }\bibfield  {title} {\bibinfo {title} {Self-consistent equations including exchange and correlation effects},\ }\href {https://doi.org/10.1103/PhysRev.140.A1133} {\bibfield  {journal} {\bibinfo  {journal} {Phys. Rev.}\ }\textbf {\bibinfo {volume} {140}},\ \bibinfo {pages} {A1133} (\bibinfo {year} {1965})}\BibitemShut {NoStop}%
\bibitem [{\citenamefont {Kresse}\ and\ \citenamefont {Furthm\"uller}(1996)}]{PhysRevB.54.11169}%
  \BibitemOpen
  \bibfield  {author} {\bibinfo {author} {\bibfnamefont {G.}~\bibnamefont {Kresse}}\ and\ \bibinfo {author} {\bibfnamefont {J.}~\bibnamefont {Furthm\"uller}},\ }\bibfield  {title} {\bibinfo {title} {Efficient iterative schemes for ab initio total-energy calculations using a plane-wave basis set},\ }\href {https://doi.org/10.1103/PhysRevB.54.11169} {\bibfield  {journal} {\bibinfo  {journal} {Phys. Rev. B}\ }\textbf {\bibinfo {volume} {54}},\ \bibinfo {pages} {11169} (\bibinfo {year} {1996})}\BibitemShut {NoStop}%
\bibitem [{\citenamefont {Kresse}\ and\ \citenamefont {Joubert}(1999)}]{PhysRevB.59.1758}%
  \BibitemOpen
  \bibfield  {author} {\bibinfo {author} {\bibfnamefont {G.}~\bibnamefont {Kresse}}\ and\ \bibinfo {author} {\bibfnamefont {D.}~\bibnamefont {Joubert}},\ }\bibfield  {title} {\bibinfo {title} {From ultrasoft pseudopotentials to the projector augmented-wave method},\ }\href {https://doi.org/10.1103/PhysRevB.59.1758} {\bibfield  {journal} {\bibinfo  {journal} {Phys. Rev. B}\ }\textbf {\bibinfo {volume} {59}},\ \bibinfo {pages} {1758} (\bibinfo {year} {1999})}\BibitemShut {NoStop}%
\bibitem [{\citenamefont {Bl\"ochl}(1994)}]{PhysRevB.50.17953}%
  \BibitemOpen
  \bibfield  {author} {\bibinfo {author} {\bibfnamefont {P.~E.}\ \bibnamefont {Bl\"ochl}},\ }\bibfield  {title} {\bibinfo {title} {Projector augmented-wave method},\ }\href {https://doi.org/10.1103/PhysRevB.50.17953} {\bibfield  {journal} {\bibinfo  {journal} {Phys. Rev. B}\ }\textbf {\bibinfo {volume} {50}},\ \bibinfo {pages} {17953} (\bibinfo {year} {1994})}\BibitemShut {NoStop}%
\bibitem [{\citenamefont {Perdew}\ \emph {et~al.}(2008)\citenamefont {Perdew}, \citenamefont {Ruzsinszky}, \citenamefont {Csonka}, \citenamefont {Vydrov}, \citenamefont {Scuseria}, \citenamefont {Constantin}, \citenamefont {Zhou},\ and\ \citenamefont {Burke}}]{PhysRevLett.100.136406}%
  \BibitemOpen
  \bibfield  {author} {\bibinfo {author} {\bibfnamefont {J.~P.}\ \bibnamefont {Perdew}}, \bibinfo {author} {\bibfnamefont {A.}~\bibnamefont {Ruzsinszky}}, \bibinfo {author} {\bibfnamefont {G.~I.}\ \bibnamefont {Csonka}}, \bibinfo {author} {\bibfnamefont {O.~A.}\ \bibnamefont {Vydrov}}, \bibinfo {author} {\bibfnamefont {G.~E.}\ \bibnamefont {Scuseria}}, \bibinfo {author} {\bibfnamefont {L.~A.}\ \bibnamefont {Constantin}}, \bibinfo {author} {\bibfnamefont {X.}~\bibnamefont {Zhou}},\ and\ \bibinfo {author} {\bibfnamefont {K.}~\bibnamefont {Burke}},\ }\bibfield  {title} {\bibinfo {title} {Restoring the density-gradient expansion for exchange in solids and surfaces},\ }\href {https://doi.org/10.1103/PhysRevLett.100.136406} {\bibfield  {journal} {\bibinfo  {journal} {Phys. Rev. Lett.}\ }\textbf {\bibinfo {volume} {100}},\ \bibinfo {pages} {136406} (\bibinfo {year} {2008})}\BibitemShut {NoStop}%
\bibitem [{\citenamefont {Monkhorst}\ and\ \citenamefont {Pack}(1976)}]{PhysRevB.13.5188}%
  \BibitemOpen
  \bibfield  {author} {\bibinfo {author} {\bibfnamefont {H.~J.}\ \bibnamefont {Monkhorst}}\ and\ \bibinfo {author} {\bibfnamefont {J.~D.}\ \bibnamefont {Pack}},\ }\bibfield  {title} {\bibinfo {title} {Special points for brillouin-zone integrations},\ }\href {https://doi.org/10.1103/PhysRevB.13.5188} {\bibfield  {journal} {\bibinfo  {journal} {Phys. Rev. B}\ }\textbf {\bibinfo {volume} {13}},\ \bibinfo {pages} {5188} (\bibinfo {year} {1976})}\BibitemShut {NoStop}%
\bibitem [{\citenamefont {Monserrat}(2018)}]{Monserrat_2018}%
  \BibitemOpen
  \bibfield  {author} {\bibinfo {author} {\bibfnamefont {B.}~\bibnamefont {Monserrat}},\ }\bibfield  {title} {\bibinfo {title} {Electron–phonon coupling from finite differences},\ }\href {https://doi.org/10.1088/1361-648X/aaa737} {\bibfield  {journal} {\bibinfo  {journal} {Journal of Physics: Condensed Matter}\ }\textbf {\bibinfo {volume} {30}},\ \bibinfo {pages} {083001} (\bibinfo {year} {2018})}\BibitemShut {NoStop}%
\bibitem [{\citenamefont {Togo}\ \emph {et~al.}(2023{\natexlab{a}})\citenamefont {Togo}, \citenamefont {Chaput}, \citenamefont {Tadano},\ and\ \citenamefont {Tanaka}}]{phonopy-phono3py-JPCM}%
  \BibitemOpen
  \bibfield  {author} {\bibinfo {author} {\bibfnamefont {A.}~\bibnamefont {Togo}}, \bibinfo {author} {\bibfnamefont {L.}~\bibnamefont {Chaput}}, \bibinfo {author} {\bibfnamefont {T.}~\bibnamefont {Tadano}},\ and\ \bibinfo {author} {\bibfnamefont {I.}~\bibnamefont {Tanaka}},\ }\bibfield  {title} {\bibinfo {title} {Implementation strategies in phonopy and phono3py},\ }\href {https://doi.org/10.1088/1361-648X/acd831} {\bibfield  {journal} {\bibinfo  {journal} {J. Phys. Condens. Matter}\ }\textbf {\bibinfo {volume} {35}},\ \bibinfo {pages} {353001} (\bibinfo {year} {2023}{\natexlab{a}})}\BibitemShut {NoStop}%
\bibitem [{\citenamefont {Carreras}\ \emph {et~al.}(2017)\citenamefont {Carreras}, \citenamefont {Togo},\ and\ \citenamefont {Tanaka}}]{Carreras_2017}%
  \BibitemOpen
  \bibfield  {author} {\bibinfo {author} {\bibfnamefont {A.}~\bibnamefont {Carreras}}, \bibinfo {author} {\bibfnamefont {A.}~\bibnamefont {Togo}},\ and\ \bibinfo {author} {\bibfnamefont {I.}~\bibnamefont {Tanaka}},\ }\bibfield  {title} {\bibinfo {title} {Dynaphopy: A code for extracting phonon quasiparticles from molecular dynamics simulations},\ }\href {https://doi.org/10.1016/j.cpc.2017.08.017} {\bibfield  {journal} {\bibinfo  {journal} {Computer Physics Communications}\ }\textbf {\bibinfo {volume} {221}},\ \bibinfo {pages} {221–234} (\bibinfo {year} {2017})}\BibitemShut {NoStop}%
\bibitem [{\citenamefont {Momma}\ and\ \citenamefont {Izumi}(2011)}]{momma_vesta_2011}%
  \BibitemOpen
  \bibfield  {author} {\bibinfo {author} {\bibfnamefont {K.}~\bibnamefont {Momma}}\ and\ \bibinfo {author} {\bibfnamefont {F.}~\bibnamefont {Izumi}},\ }\bibfield  {title} {\bibinfo {title} {{VESTA 3 for three-dimensional visualization of crystal, volumetric and morphology data}},\ }\href@noop {} {\bibfield  {journal} {\bibinfo  {journal} {J. Appl. Crystallogr.}\ }\textbf {\bibinfo {volume} {44}},\ \bibinfo {pages} {1272} (\bibinfo {year} {2011})}\BibitemShut {NoStop}%
\bibitem [{\citenamefont {Herath}\ \emph {et~al.}(2020)\citenamefont {Herath}, \citenamefont {Tavadze}, \citenamefont {He}, \citenamefont {Bousquet}, \citenamefont {Singh}, \citenamefont {Muñoz},\ and\ \citenamefont {Romero}}]{HERATH2020107080}%
  \BibitemOpen
  \bibfield  {author} {\bibinfo {author} {\bibfnamefont {U.}~\bibnamefont {Herath}}, \bibinfo {author} {\bibfnamefont {P.}~\bibnamefont {Tavadze}}, \bibinfo {author} {\bibfnamefont {X.}~\bibnamefont {He}}, \bibinfo {author} {\bibfnamefont {E.}~\bibnamefont {Bousquet}}, \bibinfo {author} {\bibfnamefont {S.}~\bibnamefont {Singh}}, \bibinfo {author} {\bibfnamefont {F.}~\bibnamefont {Muñoz}},\ and\ \bibinfo {author} {\bibfnamefont {A.~H.}\ \bibnamefont {Romero}},\ }\bibfield  {title} {\bibinfo {title} {Pyprocar: A python library for electronic structure pre/post-processing},\ }\href {https://doi.org/https://doi.org/10.1016/j.cpc.2019.107080} {\bibfield  {journal} {\bibinfo  {journal} {Computer Physics Communications}\ }\textbf {\bibinfo {volume} {251}},\ \bibinfo {pages} {107080} (\bibinfo {year} {2020})}\BibitemShut {NoStop}%
\bibitem [{\citenamefont {Lang}\ \emph {et~al.}(2024)\citenamefont {Lang}, \citenamefont {Tavadze}, \citenamefont {Tellez}, \citenamefont {Bousquet}, \citenamefont {Xu}, \citenamefont {Muñoz}, \citenamefont {Vasquez}, \citenamefont {Herath},\ and\ \citenamefont {Romero}}]{LANG2024109063}%
  \BibitemOpen
  \bibfield  {author} {\bibinfo {author} {\bibfnamefont {L.}~\bibnamefont {Lang}}, \bibinfo {author} {\bibfnamefont {P.}~\bibnamefont {Tavadze}}, \bibinfo {author} {\bibfnamefont {A.}~\bibnamefont {Tellez}}, \bibinfo {author} {\bibfnamefont {E.}~\bibnamefont {Bousquet}}, \bibinfo {author} {\bibfnamefont {H.}~\bibnamefont {Xu}}, \bibinfo {author} {\bibfnamefont {F.}~\bibnamefont {Muñoz}}, \bibinfo {author} {\bibfnamefont {N.}~\bibnamefont {Vasquez}}, \bibinfo {author} {\bibfnamefont {U.}~\bibnamefont {Herath}},\ and\ \bibinfo {author} {\bibfnamefont {A.~H.}\ \bibnamefont {Romero}},\ }\bibfield  {title} {\bibinfo {title} {Expanding pyprocar for new features, maintainability, and reliability},\ }\href {https://doi.org/https://doi.org/10.1016/j.cpc.2023.109063} {\bibfield  {journal} {\bibinfo  {journal} {Computer Physics Communications}\ }\textbf {\bibinfo {volume} {297}},\ \bibinfo {pages} {109063} (\bibinfo {year} {2024})}\BibitemShut {NoStop}%
\bibitem [{\citenamefont {Luo}\ \emph {et~al.}(2017)\citenamefont {Luo}, \citenamefont {Chen}, \citenamefont {Pei}, \citenamefont {Gao}, \citenamefont {Yan}, \citenamefont {Lu}, \citenamefont {Tong}, \citenamefont {Han}, \citenamefont {Song},\ and\ \citenamefont {Sun}}]{10.1063/1.4977708}%
  \BibitemOpen
  \bibfield  {author} {\bibinfo {author} {\bibfnamefont {X.}~\bibnamefont {Luo}}, \bibinfo {author} {\bibfnamefont {F.~C.}\ \bibnamefont {Chen}}, \bibinfo {author} {\bibfnamefont {Q.~L.}\ \bibnamefont {Pei}}, \bibinfo {author} {\bibfnamefont {J.~J.}\ \bibnamefont {Gao}}, \bibinfo {author} {\bibfnamefont {J.}~\bibnamefont {Yan}}, \bibinfo {author} {\bibfnamefont {W.~J.}\ \bibnamefont {Lu}}, \bibinfo {author} {\bibfnamefont {P.}~\bibnamefont {Tong}}, \bibinfo {author} {\bibfnamefont {Y.~Y.}\ \bibnamefont {Han}}, \bibinfo {author} {\bibfnamefont {W.~H.}\ \bibnamefont {Song}},\ and\ \bibinfo {author} {\bibfnamefont {Y.~P.}\ \bibnamefont {Sun}},\ }\bibfield  {title} {\bibinfo {title} {Resistivity plateau and large magnetoresistance in the charge density wave system tate4},\ }\href {https://doi.org/10.1063/1.4977708} {\bibfield  {journal} {\bibinfo  {journal} {Applied Physics Letters}\ }\textbf {\bibinfo {volume} {110}},\ \bibinfo {pages} {092401} (\bibinfo {year} {2017})}\BibitemShut {NoStop}%
\bibitem [{\citenamefont {B{\"o}hm}(1999)}]{Böhm1999}%
  \BibitemOpen
  \bibfield  {author} {\bibinfo {author} {\bibfnamefont {H.}~\bibnamefont {B{\"o}hm}},\ }\bibinfo {title} {X-ray crystallographic analysis of the charge-density wave modulated phases in the nbte4-tate4 system},\ in\ \href {https://doi.org/10.1007/978-94-011-4603-6_2} {\emph {\bibinfo {booktitle} {Advances in the Crystallographic and Microstructural Analysis of Charge Density Wave Modulated Crystals}}},\ \bibinfo {editor} {edited by\ \bibinfo {editor} {\bibfnamefont {F.~W.}\ \bibnamefont {Boswell}}\ and\ \bibinfo {editor} {\bibfnamefont {J.~C.}\ \bibnamefont {Bennett}}}\ (\bibinfo  {publisher} {Springer Netherlands},\ \bibinfo {address} {Dordrecht},\ \bibinfo {year} {1999})\ pp.\ \bibinfo {pages} {41--67}\BibitemShut {NoStop}%
\bibitem [{\citenamefont {Galvis}\ \emph {et~al.}(2023)\citenamefont {Galvis}, \citenamefont {Fang}, \citenamefont {Jim\'enez-Guerrero}, \citenamefont {Rojas-Castillo}, \citenamefont {Casas}, \citenamefont {Herrera}, \citenamefont {Garcia-Castro}, \citenamefont {Bousquet}, \citenamefont {Fisher}, \citenamefont {Kapitulnik},\ and\ \citenamefont {Giraldo-Gallo}}]{PhysRevB.107.045120}%
  \BibitemOpen
  \bibfield  {author} {\bibinfo {author} {\bibfnamefont {J.~A.}\ \bibnamefont {Galvis}}, \bibinfo {author} {\bibfnamefont {A.}~\bibnamefont {Fang}}, \bibinfo {author} {\bibfnamefont {D.}~\bibnamefont {Jim\'enez-Guerrero}}, \bibinfo {author} {\bibfnamefont {J.}~\bibnamefont {Rojas-Castillo}}, \bibinfo {author} {\bibfnamefont {J.}~\bibnamefont {Casas}}, \bibinfo {author} {\bibfnamefont {O.}~\bibnamefont {Herrera}}, \bibinfo {author} {\bibfnamefont {A.~C.}\ \bibnamefont {Garcia-Castro}}, \bibinfo {author} {\bibfnamefont {E.}~\bibnamefont {Bousquet}}, \bibinfo {author} {\bibfnamefont {I.~R.}\ \bibnamefont {Fisher}}, \bibinfo {author} {\bibfnamefont {A.}~\bibnamefont {Kapitulnik}},\ and\ \bibinfo {author} {\bibfnamefont {P.}~\bibnamefont {Giraldo-Gallo}},\ }\bibfield  {title} {\bibinfo {title} {Nanoscale phase-slip domain walls in the charge density wave state of the weyl semimetal candidate ${\mathrm{nbte}}_{4}$},\ }\href {https://doi.org/10.1103/PhysRevB.107.045120} {\bibfield  {journal} {\bibinfo  {journal}
  {Phys. Rev. B}\ }\textbf {\bibinfo {volume} {107}},\ \bibinfo {pages} {045120} (\bibinfo {year} {2023})}\BibitemShut {NoStop}%
\bibitem [{\citenamefont {Gruner}(1994)}]{Gruner1994}%
  \BibitemOpen
  \bibfield  {author} {\bibinfo {author} {\bibfnamefont {G.}~\bibnamefont {Gruner}},\ }\href@noop {} {\emph {\bibinfo {title} {Density Waves in Solids}}},\ Vol. 89\ (\bibinfo  {publisher} {Addison-Wesley Publishing Company},\ \bibinfo {address} {Massachusetts},\ \bibinfo {year} {1994})\BibitemShut {NoStop}%
\bibitem [{\citenamefont {Togo}\ \emph {et~al.}(2023{\natexlab{b}})\citenamefont {Togo}, \citenamefont {Chaput}, \citenamefont {Tadano},\ and\ \citenamefont {Tanaka}}]{Togo_2023}%
  \BibitemOpen
  \bibfield  {author} {\bibinfo {author} {\bibfnamefont {A.}~\bibnamefont {Togo}}, \bibinfo {author} {\bibfnamefont {L.}~\bibnamefont {Chaput}}, \bibinfo {author} {\bibfnamefont {T.}~\bibnamefont {Tadano}},\ and\ \bibinfo {author} {\bibfnamefont {I.}~\bibnamefont {Tanaka}},\ }\bibfield  {title} {\bibinfo {title} {Implementation strategies in phonopy and phono3py},\ }\href {https://doi.org/10.1088/1361-648X/acd831} {\bibfield  {journal} {\bibinfo  {journal} {Journal of Physics: Condensed Matter}\ }\textbf {\bibinfo {volume} {35}},\ \bibinfo {pages} {353001} (\bibinfo {year} {2023}{\natexlab{b}})}\BibitemShut {NoStop}%
\bibitem [{\citenamefont {Tadano}\ and\ \citenamefont {Tsuneyuki}(2015)}]{PhysRevB.92.054301}%
  \BibitemOpen
  \bibfield  {author} {\bibinfo {author} {\bibfnamefont {T.}~\bibnamefont {Tadano}}\ and\ \bibinfo {author} {\bibfnamefont {S.}~\bibnamefont {Tsuneyuki}},\ }\bibfield  {title} {\bibinfo {title} {Self-consistent phonon calculations of lattice dynamical properties in cubic ${\mathrm{srtio}}_{3}$ with first-principles anharmonic force constants},\ }\href {https://doi.org/10.1103/PhysRevB.92.054301} {\bibfield  {journal} {\bibinfo  {journal} {Phys. Rev. B}\ }\textbf {\bibinfo {volume} {92}},\ \bibinfo {pages} {054301} (\bibinfo {year} {2015})}\BibitemShut {NoStop}%
\bibitem [{\citenamefont {Horton}\ \emph {et~al.}(2025)\citenamefont {Horton}, \citenamefont {Huck}, \citenamefont {Yang}, \citenamefont {Munro}, \citenamefont {Dwaraknath}, \citenamefont {Ganose}, \citenamefont {Kingsbury}, \citenamefont {Wen}, \citenamefont {Shen}, \citenamefont {Mathis}, \citenamefont {Kaplan}, \citenamefont {Berket}, \citenamefont {Riebesell}, \citenamefont {George}, \citenamefont {Rosen}, \citenamefont {Spotte-Smith}, \citenamefont {McDermott}, \citenamefont {Cohen}, \citenamefont {Dunn}, \citenamefont {Kuner}, \citenamefont {Rignanese}, \citenamefont {Petretto}, \citenamefont {Waroquiers}, \citenamefont {Griffin}, \citenamefont {Neaton}, \citenamefont {Chrzan}, \citenamefont {Asta}, \citenamefont {Hautier}, \citenamefont {Cholia}, \citenamefont {Ceder}, \citenamefont {Ong}, \citenamefont {Jain},\ and\ \citenamefont {Persson}}]{Horton2025}%
  \BibitemOpen
  \bibfield  {author} {\bibinfo {author} {\bibfnamefont {M.~K.}\ \bibnamefont {Horton}}, \bibinfo {author} {\bibfnamefont {P.}~\bibnamefont {Huck}}, \bibinfo {author} {\bibfnamefont {R.~X.}\ \bibnamefont {Yang}}, \bibinfo {author} {\bibfnamefont {J.~M.}\ \bibnamefont {Munro}}, \bibinfo {author} {\bibfnamefont {S.}~\bibnamefont {Dwaraknath}}, \bibinfo {author} {\bibfnamefont {A.~M.}\ \bibnamefont {Ganose}}, \bibinfo {author} {\bibfnamefont {R.~S.}\ \bibnamefont {Kingsbury}}, \bibinfo {author} {\bibfnamefont {M.}~\bibnamefont {Wen}}, \bibinfo {author} {\bibfnamefont {J.~X.}\ \bibnamefont {Shen}}, \bibinfo {author} {\bibfnamefont {T.~S.}\ \bibnamefont {Mathis}}, \bibinfo {author} {\bibfnamefont {A.~D.}\ \bibnamefont {Kaplan}}, \bibinfo {author} {\bibfnamefont {K.}~\bibnamefont {Berket}}, \bibinfo {author} {\bibfnamefont {J.}~\bibnamefont {Riebesell}}, \bibinfo {author} {\bibfnamefont {J.}~\bibnamefont {George}}, \bibinfo {author} {\bibfnamefont {A.~S.}\ \bibnamefont {Rosen}}, \bibinfo {author} {\bibfnamefont
  {E.~W.~C.}\ \bibnamefont {Spotte-Smith}}, \bibinfo {author} {\bibfnamefont {M.~J.}\ \bibnamefont {McDermott}}, \bibinfo {author} {\bibfnamefont {O.~A.}\ \bibnamefont {Cohen}}, \bibinfo {author} {\bibfnamefont {A.}~\bibnamefont {Dunn}}, \bibinfo {author} {\bibfnamefont {M.~C.}\ \bibnamefont {Kuner}}, \bibinfo {author} {\bibfnamefont {G.-M.}\ \bibnamefont {Rignanese}}, \bibinfo {author} {\bibfnamefont {G.}~\bibnamefont {Petretto}}, \bibinfo {author} {\bibfnamefont {D.}~\bibnamefont {Waroquiers}}, \bibinfo {author} {\bibfnamefont {S.~M.}\ \bibnamefont {Griffin}}, \bibinfo {author} {\bibfnamefont {J.~B.}\ \bibnamefont {Neaton}}, \bibinfo {author} {\bibfnamefont {D.~C.}\ \bibnamefont {Chrzan}}, \bibinfo {author} {\bibfnamefont {M.}~\bibnamefont {Asta}}, \bibinfo {author} {\bibfnamefont {G.}~\bibnamefont {Hautier}}, \bibinfo {author} {\bibfnamefont {S.}~\bibnamefont {Cholia}}, \bibinfo {author} {\bibfnamefont {G.}~\bibnamefont {Ceder}}, \bibinfo {author} {\bibfnamefont {S.~P.}\ \bibnamefont {Ong}}, \bibinfo
  {author} {\bibfnamefont {A.}~\bibnamefont {Jain}},\ and\ \bibinfo {author} {\bibfnamefont {K.~A.}\ \bibnamefont {Persson}},\ }\bibfield  {title} {\bibinfo {title} {Accelerated data-driven materials science with the materials project},\ }\href {https://doi.org/10.1038/s41563-025-02272-0} {\bibfield  {journal} {\bibinfo  {journal} {Nature Materials}\ }\textbf {\bibinfo {volume} {24}},\ \bibinfo {pages} {1522} (\bibinfo {year} {2025})}\BibitemShut {NoStop}%
\bibitem [{\citenamefont {Nataj}\ \emph {et~al.}(2024)\citenamefont {Nataj}, \citenamefont {Kargar}, \citenamefont {Krylyuk}, \citenamefont {Debnath}, \citenamefont {Taheri}, \citenamefont {Ghosh}, \citenamefont {Zhang}, \citenamefont {Davydov}, \citenamefont {Lake},\ and\ \citenamefont {Balandin}}]{https://doi.org/10.1002/jrs.6661}%
  \BibitemOpen
  \bibfield  {author} {\bibinfo {author} {\bibfnamefont {Z.~E.}\ \bibnamefont {Nataj}}, \bibinfo {author} {\bibfnamefont {F.}~\bibnamefont {Kargar}}, \bibinfo {author} {\bibfnamefont {S.}~\bibnamefont {Krylyuk}}, \bibinfo {author} {\bibfnamefont {T.}~\bibnamefont {Debnath}}, \bibinfo {author} {\bibfnamefont {M.}~\bibnamefont {Taheri}}, \bibinfo {author} {\bibfnamefont {S.}~\bibnamefont {Ghosh}}, \bibinfo {author} {\bibfnamefont {H.}~\bibnamefont {Zhang}}, \bibinfo {author} {\bibfnamefont {A.~V.}\ \bibnamefont {Davydov}}, \bibinfo {author} {\bibfnamefont {R.~K.}\ \bibnamefont {Lake}},\ and\ \bibinfo {author} {\bibfnamefont {A.~A.}\ \bibnamefont {Balandin}},\ }\bibfield  {title} {\bibinfo {title} {Raman spectroscopy of phonon states in nbte4 and tate4 quasi-one-dimensional van der waals crystals},\ }\href {https://doi.org/https://doi.org/10.1002/jrs.6661} {\bibfield  {journal} {\bibinfo  {journal} {Journal of Raman Spectroscopy}\ }\textbf {\bibinfo {volume} {55}},\ \bibinfo {pages} {695} (\bibinfo {year}
  {2024})}\BibitemShut {NoStop}%
\end{thebibliography}%
\end{document}